\documentstyle[prb,twocolumn,aps,graphicx,eqsecnum]{revtex}
\begin{document}
\draft

 \newcommand{\mytitle}[1]{
 \twocolumn[\hsize\textwidth\columnwidth\hsize
 \csname@twocolumnfalse\endcsname #1 \vspace{1mm}]}

\mytitle{
\title{Aharonov-Bohm Interferometry with Interacting Quantum Dots:
Spin Configurations, Asymmetric Interference Patterns, 
Bias-Voltage-Induced Aharonov-Bohm Oscillations, and 
Symmetries of Transport Coefficients}

\author{J\"urgen K\"onig$^{1,2}$ and Yuval Gefen$^{3}$}
\address{
$^1$Institut f\"ur Theoretische Festk\"orperphysik, Universit\"at Karlsruhe,
76128 Karlsruhe, Germany\\
$^2$Department of Physics, The University of Texas at Austin, Austin, Texas
78712, USA\\
$^3$Department of Condensed Matter Physics, The Weizmann Institute of Science,
76100 Rehovot, Israel}

\date{\today}

\maketitle

\begin{abstract}

We study electron transport through multiply connected mesoscopic geometries
containing interacting quantum dots.
Our formulation covers both equilibrium and nonequilibrium physics.
We discuss the relation of coherent transport channels through the quantum dot
to flux-sensitive Aharonov-Bohm oscillations in the total conductance of the
device.
Contributions to transport in first and second order in the intrinsic 
linewidth of the dot levels are addressed in detail.
We predict an interaction-induced asymmetry in the amplitude of the
interference signal around resonance peaks as a consequence of incoherence
associated with spin-flip processes.
This asymmetry can be used to probe the total spin of the quantum dot.
Such a probe requires less stringent experimental conditions than the Kondo
effect, which provides the same information.
We show that first-order contributions can be partially or even fully coherent.
This contrasts with the sequential-tunneling picture, which describes
first-order transport as a sequence of incoherent tunneling processes.
We predict bias-voltage-induced Aharonov-Bohm oscillations of physical
quantities which are independent of flux in the linear-response regime.
Going beyond the Onsager relations we analyze the relations between the space
symmetry group of the setup and the flux-dependent nonlinear conductance.

\end{abstract}
}

%
%

\section{Introduction}
The manifestation of quantum coherence in finite systems is in the foundations
of the physics of mesoscopic systems.
The presence of quantum coherence is detectable through interference
experiments, most notably Aharonov-Bohm (AB) interferometry.
The prototype of an AB setup is a double-slit experiment, as shown in
Fig.~\ref{fig1}a.
An electron moving from the left reservoir to the right one is split into two
partial waves which interfere with each other.
A magnetic flux $\Phi$ that penetrates the area enclosed by the two paths
changes the relative phase of the amplitudes of the two partial waves, which
yields a flux dependence of the total transmission probability through the
device.

A convenient framework to study the role of electron-electron (e-e)
interactions in mesoscopic systems is suggested by employing quantum dots (QDs).
The latter provide a relatively simple and controlled scheme to address e-e
interactions without undermining the rich physics involved.
Transport through QDs has been studied extensively and revealed interesting
phenomena such as resonant tunneling, Coulomb blockade, and the Kondo effect.
However, the measurement of the current through QDs does not provide
information about the quantum coherence of the transport.
In particular, it doesn't address the interplay of electron-electron
interactions and quantum coherence.
In order to approach these questions QDs have been embedded in AB geometries.%
\cite{Akera93,Yacobi95,Yeyati95,Hackenbroich96,Bruder96,Oreg97,Schuster97,%
Izumida97,Baltin99.1,Baltin99.2,Ji00,Wiel00,Loss00,Gerland00,%
Silvestrov00,Kang00,Boese00,Holleitner00,Koenig01,Hofstetter01}
Magnetic-flux sensitivity of the total current has been observed,%
\cite{Yacobi95,Schuster97,Ji00,Wiel00,Holleitner00} indicating at least
partially coherent transport through the QD.

By applying a finite bias voltage across the device the QD can be driven out
of equilibrium, lifting the complexity of the system to a qualitatively higher
level.
In the context of interacting QDs nonequilibrium effects such as splitting of
Kondo resonances have been discussed.
It is only natural to expect that driving the system out of equilibrium
enriches the interplay between quantum coherence and interaction effects by
another facet.
Some of the symmetries present at equilibrium, which underline linear response,
may be broken, and at the same time new qualitative features may emerge.

In this paper we address two main issues:

(i) What can we learn about QDs by embedding them into AB geometries?
Based on Ref.~\onlinecite{Koenig01}, we will classify different contributions
to linear transport into coherent and incoherent channels.
We concentrate on the dominant transport mechanism, which is of first order in
the intrinsic linewidth of the dot levels when the QD is tuned in resonance
with the leads and of second order (``cotunneling'') in the Coulomb blockade
regime.\cite{note1}
A sufficient condition to establish {\it full} coherence is to find an AB
geometry in which the interference pattern reveals {\it fully destructive
interference} since transport processes through the QD which are
{\it cancelled} by {\it adding} parallel transmission channels cannot be
incoherent.
As we will see the proper choice of the AB interferometer, including either one
or two QDs, is important for the manifestation of this criterion.

A very important and striking result of our analysis will lead to the
suggestion that AB interferometers (with either one or two QDs) can be used to
probe the total spin of the QD.
We will find that the relative magnitude of the flux-dependent AB signal, the
visibility, changes from one Coulomb blockade valley to the next.
This amounts to an asymmetry in the AB signal around the Coulomb conductance
peaks.
The higher visibility indicates the side of the peak which corresponds to a
situation where the QD is mostly occupied by an even number of electrons
(doubly occupied levels) with a total spin $0$.
The lower visibility corresponds to an odd number of electrons and total
spin $1/2$.
A similar even/odd effect in the low-temperature conductance through a QD
results from the Kondo effect.\cite{Goldhaber98,Cronenwett98,Simmel99,%
Schmid00,Glazman88,Ng88,Meir93,Koenig96}
Thus, measurement of the conductance probes the total spin of the QD, too,
although the physical origin is completely different.
We stress that our spin-dependent asymmetry effect is {\it not} a manifestation
of incipient Kondo physics.
Indeed, we expect our asymmetry to disappear in the zero-temperature limit,
where the Kondo effect is fully developed.
In order to access the Kondo regime, low temperature and strong
coupling of the dot to the leads is required.
This contrasts with our suggestion to employ an AB interferometer.
We predict the asymmetry of the AB signal to survive at high temperature and
weak coupling, i.e., much less stringent conditions.

(ii) The other major issue of this paper addresses the interplay between
quantum coherence and electron-electron interaction for systems
{\it out of equilibrium}.
In the linear-response regime the average electron number on the QD is an
equilibrium quantity.
For weak coupling between dot and leads (such that the intrinsic linewidth is
small compared to the energy scale provided by the temperature and the level
spacing), the dot occupation is governed by classical Boltzmann factors, which
are independent of magnetic flux.
Out of equilibrium, however, we predict {\it bias-voltage-induced AB
oscillations} in the dot occupation for a single-dot AB interferometer.
We emphasize that, again, the experimental conditions for this effect to occur,
high temperature and weak coupling, are accessible rather easily.

Qualitative differences between equilibrium and nonequilibrium situations
also show up in symmetry relations of transport coefficients.
The classical double-slit setup (Fig.~\ref{fig1}a) is an example of an
{\it open-geometry} interferometer, where those electrons which are not
transmitted directly to the lead on the right are absorbed by other gates
and terminals of the system at the periphery of the device (these are not
shown in the figure).
In most parts of the paper, however, we consider AB interferometers in a
{\it closed geometry}, as sketched in Fig.~\ref{fig1}b, where these electrons
can only go to either the terminal on the left or the one on the
right-hand-side.
For the open-geometry setup (Fig.~\ref{fig1}a) $\Phi=0$ is {\it not} a
symmetry point (i.e., the transmission, as well as transport coefficients and
the partition function, are not invariant under $\Phi \rightarrow -\Phi$).
The symmetry point depends on specifics of the interferometer.
This contrasts with the case of a two-terminal geometry (Fig.~\ref{fig1}b),
where all thermodynamic potentials as well as {\it linear-response} transport
coefficients are invariant under the inversion of the flux.\cite{Buettiker86}
The fact that $\Phi=0$ is an extremum point is usually referred to as
{\it phase locking}, and is a direct consequence of Onsager relations.
It has been demonstrated\cite{Bruder96} that beyond linear response, i.e., in
the presence of a finite voltage bias, phase locking is, in general, broken.
We show by an explicit calculation that for a two-terminal AB interferometer
including a single QD {\it both} electron-electron interaction {\it and} finite
bias voltage is needed to break phase locking.

But even in the nonlinear-response regime, there are symmetry relations in the
transport coefficients which are connected to the space-symmetry group of
the setup.
This will be especially important for AB interferometers with two QDs, where,
depending on the spatial symmetry, phase locking can occur even at finite
bias voltages.
\begin{figure}
\centerline{\includegraphics[width=8cm]{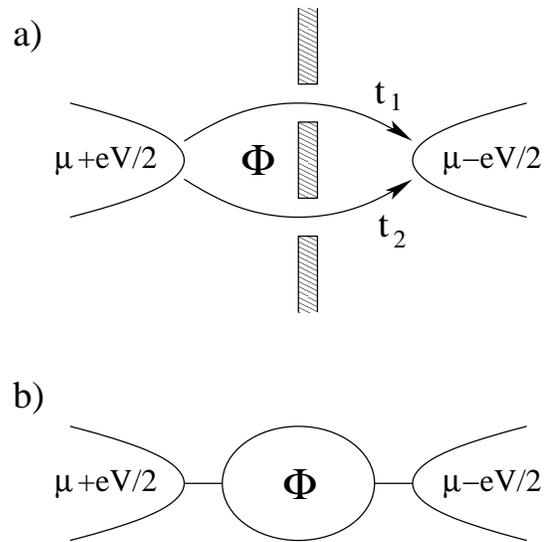}}
\caption{a) Two-slit experiment as an open-geometry AB interferometer: only
        a fraction of the electrons passing through the interferometer will
        reach the drain.
        b) Two-terminal (closed-geometry) AB interferometer: all electrons
        injected from the source must either reach the drain or be reflected to
        the source.
        }
\label{fig1}
\end{figure}
We conclude this introduction with some remarks about the formalism to be
employed.
The experimentally accessible quantity which characterizes transport is the
total current through the device, which is driven by bias voltage $V$.
The electrical current (flowing from the right to the left) can be written in
the form
\begin{equation}
\label{nonlinear_current}
   I = {e\over h}\sum_\sigma \int d \omega \, T_\sigma(\omega)
        [f_L(\omega) - f_R(\omega)]  \, ,
\end{equation}
where $f_r(\omega) = 1/[1+\exp(\beta(\omega-\mu_r))]$ with $r=L/R$ is the
Fermi-Dirac distribution function, and $\beta = 1/(k_BT)$ is the inverse
temperature.
The left and the right lead have fixed electrochemical potentials
$\mu_L = \mu + eV/2$ and $\mu_R = \mu - eV/2$ for the electrons, respectively,
with $e>0$.
The quantity $T_\sigma(\omega)$ defines the probability for an incoming
electron with energy $\omega$ (measured with respect to $\mu$) and spin
$\sigma$ to be transmitted through the device.

According to the Landauer-B\"uttiker or scattering
formalism,\cite{Landauer70,Buettiker85} which we are not going to use for
reasons explained below, the transmission {\it probability} $T_\sigma(\omega)$
for each channel labeled by $\omega$ and $\sigma$ is obtained from the modulus
squared of the total transmission {\it amplitude}.
In the example of the double-slit setup, Fig.~\ref{fig1}a, we have
$T_\sigma(\omega) = |t_{1\sigma}(\omega)\,+\,t_{2\sigma}(\omega)|^2$, where
$t_1$ and $t_2$ are the partial amplitudes through either slit and are complex
quantities.
The associated phases have an orbital contribution, determined by the
geometrical details of the interferometer, and a magnetic-flux-dependent part.
The transmission probability is, then, given by
\begin{equation}
\label{transmission_phases}
   T = |t_1|^2 \, + \, |t_2|^2 \, +
        2|t_1t_2| \cos (\varphi + \delta \theta) \, ,
\end{equation}
where $\delta \theta$ is the relative orbital phase,
$\varphi \equiv 2\pi \Phi / \Phi_0$ accounts for the enclosed magnet flux, and
$\Phi_0 = h/e$ is the flux quantum.
(Here, for this geometry, it is justified to neglect the multiply scattered
higher-winding-number amplitudes.
For closed geometries, such as in Fig.~\ref{fig1}b, the partial amplitudes
may exhibit a more complicated dependence on magnetic field, representing
partial amplitudes with higher winding numbers around the enclosed flux.)
The success of the Landauer-B\"uttiker picture lies in part in the fact that,
once the partial transmission amplitudes are known, the analysis is
straightforward.
Indeed, for {\it noninteracting} systems, the transmission amplitude for
an electron to travel from the left to the right lead is simply determined from
the (retarded) single-particle Green's function $G^r_{LR,\sigma} (\omega)$
which involves a creation operator of an electron in the right and an
annihilation operator in the left lead.
{\it This single-particle picture, however, breaks down if electron-electron
interaction is taken into account.}
This has been shown in the literature\cite{Meir92} and will be explicitly
demonstrated in the next section.
In this case, we do not see a general recipe of how to construct meaningful
transmission amplitudes.

Instead we directly calculate the current using Green's function techniques
for {\it interacting} systems,\cite{Koenig96,Meir92,Koenig99} and extract
the transmission {\it probability} from Eq.~(\ref{nonlinear_current}) without
worrying about transmission amplitudes.

The outline of this paper is as follows.
In Sec.~\ref{section_general} we employ a simple physical picture to
distinguish coherent from incoherent transport through a QD.
To substantiate these qualitative considerations we calculate in
Sec.~\ref{section_one_dot} the current through an AB interferometer in which a
single QD in embedded.
Interferometry with two QDs is addressed in Sec.~\ref{section_two_dots}.
Section~\ref{section_finite_bias} is devoted to nonequilibrium physics, and
in Sec.~\ref{section_symmetry} we draw the connection between the spatial
symmetry of the setup and symmetry relations of nonlinear transport
coefficients.
We have also included a few appendices to provide some technical
details of our analysis.
In Appendix~\ref{append_one_dot} and \ref{append_two_dots} we determine the
density matrix for the single-dot and two-dot AB interferometer, respectively,
and in Appendix~\ref{append_alternative} we present an alternative derivation of
Eq.~(\ref{two dots noninteracting}).

The main results of our analysis  are summarized in 
(i) Eqs.~(\ref{current_0}) and (\ref{current_1}) (the zeroth and first-order 
expressions for the current through the interferometer); 
(ii) Eqs.~(\ref{trans_1a}) and (\ref{trans_1b}) (the flux-dependent 
linear-response transmission for the closed and open geometry respectively); 
(iii) Eqs.~(\ref{cotunneling noninteracting}), (\ref{cotunneling interacting}),
(\ref{two dots noninteracting}), (\ref{two dots interacting}) (the
absence/presence of asymmetry in the case of $U=0$, $U=\infty$ respectively),
for the single QD in the cotunneling regime, and the double QD at resonance;
(iv) the various regimes for the multilevel dot indicated at the end of
Section~\ref{subsection_many_levels}; 
(v) Eqs.~(\ref{induced oscillation one dot}) and 
(\ref{induced oscillation two dots})
(the finite-bias induced AB oscillations in the dot occupation 
$\langle N\rangle$); 
(vi) Eqs.~(\ref{symmetry 1}) - (\ref{symmetry 3}) (the relation between 
flux-reversal symmetry and space-symmetry, away from linear response).

\section{Coherent versus incoherent transport through a QD}
\label{section_general}

Before analyzing AB interferometers which contain QDs, as shown in 
Figs.~\ref{fig2}b and c, we begin with considering transport through a QD alone, 
Fig.~\ref{fig2}a.
\begin{figure}
\centerline{\includegraphics[width=8cm]{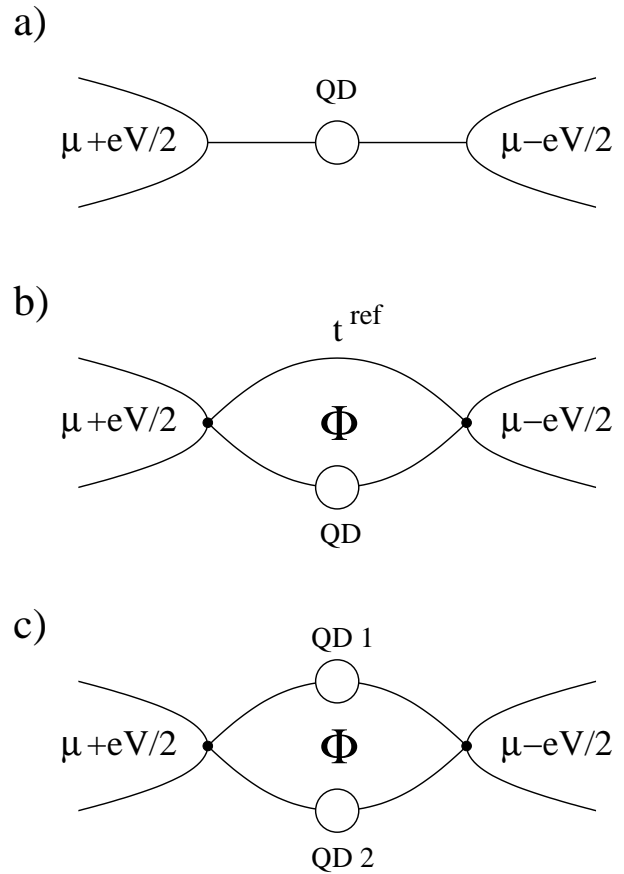}}
\caption{Transport through a QD and AB interferometers with either one or two
        QDs.}
\label{fig2}
\end{figure}
We develop a simple physical picture to underline the essential physics
involved in the predicted interaction-induced asymmetry of the interference
signal around resonance peaks.
All statements made in this section will be backed up by strict calculations
for the AB interferometers in the subsequent sections.
Furthermore, we give an explicit example of how the single-particle formalism
breaks down in the presence of Coulomb interaction.

We consider a single-level QD with level energy $\epsilon$, measured from the
Fermi energy of the leads.
The Hamiltonian
\begin{equation}
\label{Hamiltonian}
   H^{\rm dot}=H_L+H_R+H_D+H_T
\end{equation}
contains a part describing noninteracting electrons in the left and right
leads,
\begin{equation}
   H_r = \sum_{k \sigma} \epsilon_{kr} a^\dagger_{k\sigma r}a_{k\sigma r}
\end{equation}
with $r=L/R$.
The isolated dot is described by
\begin{equation}
   H_D = \epsilon \sum_\sigma n_\sigma + U n_\uparrow n_\downarrow \,\, ,
\end{equation}
where $n_\sigma = c^\dagger_\sigma c_\sigma$ counts the number of electrons
with spin $\sigma$.
The energy of the dot level $\epsilon$ can be varied by an applied gate
voltage.
The electron-electron interaction is accounted for by the charging energy
penalty $U=2(e^2/(2C))$ for double occupancy, where $C$ is the effective
capacitance of the QD.
To keep the discussion simple we choose the generic limits $U=0$ for the
noninteracting case and $U=\infty$ for an interacting QD.
We stress that the pursuant analysis, and in particular our qualitative
conclusions, are applicable to the case of finite $U$ as well.
Tunneling between the QD and the leads is modeled by
\begin{equation}
   H_T = \sum_{k \sigma r} (t_r a^\dagger_{k\sigma r} c_\sigma + {\rm H.c.})
        \, .
\end{equation}
(We neglect the energy dependence of the tunnel matrix elements $t_{L/R}$).
Due to tunneling the dot level acquires a finite linewidth
$\Gamma=\Gamma_L+\Gamma_R$ with
\begin{equation}
\label{gamma}
   \Gamma_{L/R}=2\pi|t_{L/R}|^2N_{L/R} \,\, ,
\end{equation}
where $N_{L/R}$ is the density of states in the corresponding lead.

For the sake of developing physical insight, we consider an off-resonance
scenario, i.e., the gate voltage is tuned in such a way that
$|\epsilon| \gg k_BT, \Gamma$.
In this regime (and as long as the temperature is higher than the Kondo
temperature), transport is dominated by second-order processes in $\Gamma$,
which are usually called cotunneling.\cite{Averin89,renormalization}
Furthermore, we assume an infinite large charging energy $U=\infty$.
Figure~\ref{fig3} shows all processes (second order in $\Gamma$) which contribute
to the current from left to right in which the incoming electron has spin up.
Similarly, there are processes in which the incoming electron has spin down.
They will introduce a trivial factor of 2 to the final expressions for the
current.
The three possible second-order processes are\\
(a) a (spin-up) electron enters the QD, leading to a virtual occupancy, and
then leaves it to the other side.\\
(b) a (spin-up) electron leaves the QD, and an electron with the
same spin enters.\\
(c) a (spin-down) electron leaves the QD, and an electron with opposite spin
enters.
\begin{figure}
\centerline{\includegraphics[width=8cm]{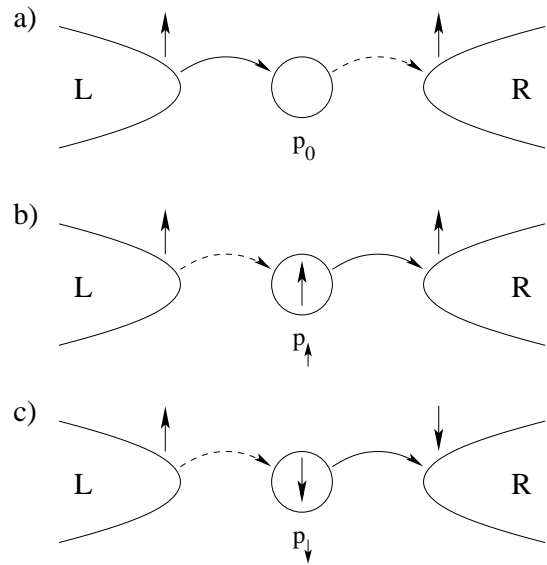}}
\caption{Cotunneling processes for $U=\infty$.
        The solid line indicates the process that happens first, the dashed
        line the process that occurs afterwards.
        Double occupancy of the QD in the initial, intermediate, or final
        state is prohibited by charging energy.}
\label{fig3}
\end{figure}
Note that double occupancy of the dot (even in a virtual state) is forbidden
since we have assumed $U=\infty$.
All processes are elastic in the sense that the energy of the QD has not
changed between its initial and final states.
Process (c), though, is incoherent since the spin in the the QD has been
flipped.
In other words, the traversing electron has left a trace in the ``environment''
(i.e., the dot).
This implies that when the QD is embedded in one of the arms of an AB
interferometer, one will be able to tell that an electron participating in
process (c) has indeed traversed the arm with the QD (rather than the other
arm).
Such a process will then contribute to the total current but not to the
flux-sensitive component thereof, independent of the realization of the AB
interferometer.
The notion that energy exchange is not necessary for dephasing,%
\cite{Altshuler82}
and that the latter can take place through, e.g., a spin flip of an external
degree of freedom, has been made early on.\cite{Stern90}
In our case the electrons in the QD itself (and their spin) serve as the
``dephasing bath''.\cite{Mello00}

Our simple picture already indicates that the physics in adjacent Coulomb
blockade valleys is qualitatively different.
This is the origin of the asymmetry in the visibility discussed throughout
this paper.
In our model there are two possible charge states for the QD: the dot
being empty, $N=0$, or singly occupied, $N=1$, with either spin up or down.
By tuning the gate voltage (which changes the level position $\epsilon$) one
can drive the system away from resonance, where both states $N=0$ and $N=1$
are possible, to the Coulomb blockade valley with a fixed electron number.
The latter is either $N=0$ or, on the other side of the resonance, $N=1$.
In the first case, the main contribution to the current is due to the
coherent process (a) in Fig.~\ref{fig3}.
In the latter case, two processes, (b) and (c), contribute to the current but
only one of them, namely (b), is coherent.
We thus expect that in an AB interferometer with a single QD the AB amplitude
at a gate voltage $V_g^{(0)}$ in the $N=0$ valley will be twice as large as
the AB amplitude at a gate voltage $V_g^{(1)}$ in the $N=1$ valley.
For such a comparison one needs to consider gate voltages $V_g^{(0)}$ and
$V_g^{(1)}$ for which the flux-averaged transmission is the same.
We also note that a multilevel system may offer more complex configurations,
e.g., a resonance which separates a valley with all levels doubly occupied from
a valley where one extra electron is added and one, three, or more levels are now
singly occupied.

The transmission probabilities of electrons with energy $\omega$ near the
Fermi level of the leads can be obtained by calculating the transition rate in
second-order perturbation theory and multiplying it with the probabilities
$P_\chi$ to find the system in the corresponding initial state $\chi$.
For an incoming electron with spin up the transmission probabilities are
$P_\chi \Gamma_L \Gamma_R {\rm Re}[1/(\omega-\epsilon+i0^+)^2]$ with
$\chi=0,\uparrow,\downarrow$ for case (a), (b), and (c), respectively.
Since $P_0 +P_\uparrow +P_\downarrow =1$ and $P_0+P_\sigma =1/[1+f(\epsilon)]$
in equilibrium, where $f(\epsilon)$ is the Fermi function, we find for the
transmission probability through the QD that
$T^{\rm dot}_\sigma(\omega) = T^{\rm dot, coh}_\sigma(\omega) +
T^{\rm dot, incoh}_\sigma(\omega)$ with\cite{comment_reg}
\begin{eqnarray}
\label{cotunneling total}
   T^{\rm dot}_\sigma(\omega) &=& {\rm Re} \, {\Gamma_L \Gamma_R \over
        (\omega - \epsilon + i0^+)^2} \, ,
\\
\label{cotunneling coherent}
   T^{\rm dot, coh}_\sigma(\omega) &=& {T^{\rm dot}_\sigma(\omega)\over
        1+f(\epsilon)} \, .
\end{eqnarray}
In linear response the participating electrons have energies $\omega$ that are
spread by $k_B T$ around the Fermi energy.
Since we consider an off-resonance situation, we can set $\omega \approx 0$ in
Eqs.~(\ref{cotunneling total}) and (\ref{cotunneling coherent}) and find that,
while the total transmission is symmetric under
$\epsilon \rightarrow -\epsilon$, the factor $1/[1+f(\epsilon)]$ ascribes an
asymmetry to the coherent part.

We have argued that incoherence is induced by flipping the spin of the
transfered electron, Fig.~\ref{fig3}c.
One might expect naively that such spin-flip processes may take place for
noninteracting systems as well.
On the other hand, in the absence of interaction, transport should be fully
coherent.
This puzzle is solved by the observation that in the absence of interaction,
$U=0$, double occupancy of the dot is allowed and, therefore, more processes
are possible.
Figure~\ref{fig4} depicts these additional second-order processes (with an
incoming spin-up electron).
Spin flip takes place in Figs.~\ref{fig3}c and \ref{fig4}c only.
The first is hole-like while the second is particle-like (the solid line
indicates the first step and the dashed line the second step of the process).
Both amplitudes have the same magnitude but come with opposite signs and,
hence, cancel out.
\begin{figure}
\centerline{\includegraphics[width=8cm]{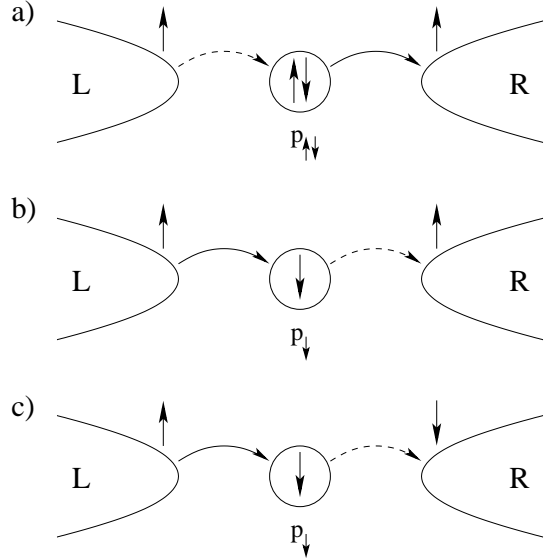}}
\caption{Additional cotunneling processes for $U < \infty$.
        Now also double occupancy of the QD is possible.}
\label{fig4}
\end{figure}
Thus, {\it in the absence of Coulomb interaction} (and no coupling to an
external environment) {\it there are no spin-flip processes, and the
transmission is fully coherent}.
We perform an analogous second-order perturbation calculation as above, use
$P_0 + P_\uparrow + P_\downarrow + P_{\uparrow\downarrow} = 1$, and find that
$T^{\rm dot}_\sigma(\omega) = T^{\rm dot, coh}_\sigma(\omega)$ given by
Eq.~(\ref{cotunneling total}) without any asymmetry factor.
Once $U$ is turned on, the exact cancellation of the spin-flip processes is
lost, and incoherent transmission channels induce an asymmetry of the AB
signal.

As mentioned in the previous section, for noninteracting systems the
transmission amplitude is determined by the retarded single-particle Green's
function from the left to the right lead, and then we can easily apply the
Landauer-B\"uttiker formalism.
In our case this amplitude can be expressed as
\begin{equation}
\label{amplitude}
   t_\sigma^{\rm dot}(\omega) = i \sqrt{\Gamma_L \Gamma_R} \,
        G^r_\sigma (\omega) \, ,
\end{equation}
where the dot Green's function $G^r_\sigma (\omega)$ is the Fourier transform
of $-i\Theta (t) \langle \left\{ c_\sigma (t),c^\dagger_\sigma (0) \right\}
\rangle$.
We insert the dot Green's function (for $U=0$ and zeroth order in $\Gamma$)
into Eq.~(\ref{amplitude}) and get $t_\sigma^{\rm dot}(\omega) =
i \sqrt{\Gamma_L \Gamma_R} / (\omega-\epsilon + i0^+)$ which yields
\begin{equation}
   |t_\sigma^{\rm dot}(\omega)|^2 = T_\sigma^{\rm dot}(\omega) \, ,
\end{equation}
with $T_\sigma^{\rm dot}(\omega)$ given by Eq.~(\ref{cotunneling total}).

{\it In the presence of interaction, however, Eq.~(\ref{amplitude}) is no
longer a good definition for the transmission amplitude.}
To see this we calculate the dot Green's function in the limit $U=\infty$
and zeroth order in $\Gamma$.
We find $t_\sigma^{\rm dot}(\omega) = i (P_0 +P_\uparrow) \sqrt{\Gamma_L
\Gamma_R} / (\omega-\epsilon + i0^+)$ and, therefore,
\begin{equation}
\label{amplitude squared}
   |t_\sigma^{\rm dot}(\omega)|^2 = {T_\sigma^{\rm dot}(\omega) \over
        [1+f(\epsilon)]^2} \, .
\end{equation}
We see that not only does $|t_\sigma^{\rm dot}(\omega)|^2$ not yield the total
transmission Eq.~(\ref{cotunneling total}) through the dot, but it differs from
the coherent part of the transmission, Eq.~(\ref{cotunneling coherent}),
as well: there is no direct physical meaning of the expression
$|t_\sigma^{\rm dot}(\omega)|^2$.

Can we find a similar clear picture for first-order transport in $\Gamma$,
which dominates at resonance?
As long as quantum coherence is not addressed, first-order transport through
QDs has been successfully described by the use of the sequential-tunneling
picture or ``orthodox theory''.
Tunneling rates into and out of the QD are calculated with Fermi's golden rule.
They are connected to the probabilities for the dot states by a master
equation, and determine the current.
According to this philosophy, different tunneling processes in or out of the QD
are uncorrelated since it is implicitly assumed that the electron experiences
some dephasing during its stay in the QD.
As a consequence the sequential-tunneling picture implies that first-order
transport should be completely incoherent.
This may be well justified for large QDs which are coupled to phonons or other
degrees of freedom.
But for the system under consideration in this paper, a single-level QD with no
coupling to external baths, this is not the case.
For $U=0$ we even expect fully coherent transport to all orders in $\Gamma$.
As we will prove in Sec.~\ref{section_two_dots} by strict calculations, the
sequential-tunneling picture completely misjudges the coherence of first-order
transport, although it predicts the correct transmission probability through
a QD.
In our case, first-order transport should rather be viewed as ``resonant
tunneling'' expanded up to first order than visualized as a sequence of
(incoherent) sequential-tunneling processes.

\section{Interferometry with a single QD}
\label{section_one_dot}

To support the results of our intuitive picture, we analyze quantitatively
AB interferometers which contain a single QD, such as shown in Fig.~\ref{fig2}b.
We emphasize that all following results (except for the discussion of many
levels in Sec.~\ref{subsection_many_levels}) are derived by strict calculation
of the total current through the device in the presence of magnetic flux.
At no point we make use of the qualitative picture outlined in the previous
section.
The compatibility of our calculations with the intuitive picture serves rather
as a check.

Electrons emitted from the left lead have two possible ways to reach the
drain on the right.
They can either go through the QD in the lower arm or choose the upper arm.
We denote the transmission amplitude for the latter trajectory by
$t^{\rm ref}$ and assume that it is independent of energy $\omega$ and spin
$\sigma$.
The total transmission probability $T^{\rm tot}_\sigma(\omega)$ through this
device is the sum of three parts,
\begin{equation}
    T^{\rm tot}_\sigma(\omega) = T^{\rm dot}_\sigma(\omega)
        + T^{\rm ref}_\sigma + T^{\rm flux}_\sigma (\omega) \, .
\end{equation}
The terms $T^{\rm dot}_\sigma(\omega)$ and $T^{\rm ref}_\sigma = |t^{\rm ref}|^2$
are the flux-insensitive transmissions through the QD and the reference arm,
respectively.
(In principle, $T^{\rm dot}_\sigma(\omega)$ feels the presence of the reference
arm and vice versa.
When low-order transmission in the couplings is considered this influence may
be ignored.)
Interference is described by the remaining term,
$T^{\rm flux}_\sigma (\omega)$, which depends on magnetic flux.

\subsection{Current formula}

In the first step we relate the flux-dependent nonlinear current to the dot
Green's function for the closed geometry, Fig.~\ref{fig2}b.
This derivation is somewhat technical but straightforward.
The central result of this part is the current in first order in $t^{\rm ref}$,
Eq.~(\ref{current_1}) [along with the zeroth-order result,
Eqs.~(\ref{current_0})].
The total Hamiltonian,
\begin{equation}
   H = H^{\rm dot} + H^{\rm ref} \, ,
\end{equation}
consists of two parts.
The first, $H^{\rm dot}$, given by Eq.~(\ref{Hamiltonian}), describes the
arm containing the QD, and the second,
\begin{equation}
\label{H_ref}
   H^{\rm ref} = \sum_{k\in R,q\in L,\sigma} (\tilde t
        a^\dagger_{k\sigma R} a_{q\sigma L} + {\rm H.c.}) \, ,
\end{equation}
with $2\pi \tilde t \sqrt{N_L N_R} = |t^{\rm ref}| e^{i\varphi}$, models the
transmission through the reference arm.
In general, the magnetic flux $\Phi$ enters the phases of all three tunneling
matrix elements $t_L$, $t_R$, and $t^{\rm ref}$.
Above, however, we have chosen a gauge in which only $\tilde t$ acquires a
flux-dependent phase $\varphi = 2 \pi \Phi /\Phi_0$ but leaves $t_L$ and $t_R$
flux independent, and we can choose the latter to be real.

The operator for the current from the right lead is given by the time
derivative of the total electron number operator $\hat n_R =
\sum_{k \in R,\sigma} a^\dagger_{k\sigma R} a_{k\sigma R}$ times the elementary
charge $e$.
This yields for the total current
\begin{equation}
    I = I_R = e {d \langle \hat n_R \rangle \over dt} = i{e\over \hbar}
        \langle [ \hat H, \hat n_R ] \rangle \, .
\end{equation}
The latter expression yields Green's functions which involve Fermi operators
of the right lead and of either the left lead or the dot,
\begin{eqnarray}
   I_R &=& -{e\over h}
        \sum_{q\in L,k\in R,\sigma} \int d \omega \left[
        \tilde t \, G^<_{qk,\sigma}(\omega) + {\rm H.c.} \right]
\nonumber \\
        && - {e\over h}\sum_{k\in R,\sigma} \int d \omega \left[
        t_R \, G^<_{dk,\sigma}(\omega) + {\rm H.c.} \right] \, ,
\label{current_start}
\end{eqnarray}
with the notations
$G^<_{qk,\sigma}(t) = i \langle a^\dagger_{k\sigma R}(0) a_{q\sigma L}(t)\rangle$
and
$G^<_{dk,\sigma}(t) = i \langle a^\dagger_{k\sigma R}(0) c_{\sigma}(t)\rangle$.
The indices $q$ and $k$ label the states in the left and right leads,
respectively.
The index $d$ indicates that a dot electron operator is involved (in our
simple model there is only one dot level).
The other Green's functions are defined similarly,
$G^<_{kq,\sigma}(t) = i \langle a^\dagger_{q\sigma L}(0) a_{k\sigma R}(t)\rangle$
and
$G^<_{kd,\sigma}(t) = i \langle c_{\sigma}^\dagger(0) a_{k\sigma R}(t)\rangle$.
Note that, since the $z$ component of the spin of the entire system is conserved
(as well as the total spin), only Green's functions which are diagonal in spin
space are involved.

The first (second) line of Eq.~(\ref{current_start}) describes electron
transfer from the left (from the QD) to the right lead or vice versa.
This transfer can be a direct tunneling process or a complex trajectory
through the entire device.
We emphasize that Eq.~(\ref{current_start}) is exact as long as the full
Green's functions with contributions of arbitrary high order in $t$ and
$t^{\rm ref}$ are inserted.

Our goal is to derive a relation between the current and Green's functions
involving only dot operators.
To achieve this we employ the Keldysh technique and use the matrix
representation
\begin{equation}
   {\bf G} = \left( \begin{array}{cc} G^r & G^< \\ 0 & G^a \end{array} \right)
        \, ,
\end{equation}
where $G^r$ and $G^a$ are the usual retarded and advanced Green's functions,
respectively.

The contribution to the total current can be classified by powers of
$t^{\rm ref}$.
The zeroth-order term $I_R^{(0)}$ represents the current through the QD in
the absence of the reference arm and is independent of magnetic flux.
The main contribution $I_R^{(1)}$ to the flux-dependent part is first order in
$t^{\rm ref}$.
In the spirit of a series expansion we drop in the flux-dependent part all
second- or higher-order terms of $t^{\rm ref}$, which is a good approximation as
long as the transmission through the reference arm is small (trajectories with
higher winding numbers, where an electron moves around the enclosed flux several
times, are described by such higher-order terms).
Transmission through the reference arm, which is at least of order
$(t^{\rm ref})^2$, is not considered in this subsection.

We start with the zeroth-order term, $I_R^{(0)}$.
Only the second line of Eq.~(\ref{current_start}) has to be included since the
first line explicitly contains $\tilde t$.
Since the electrons in the leads are noninteracting, we get the Dyson-like
equation ${\bf G}^{(0)}_{dk,\sigma} (\omega) =
{\bf G}_\sigma^{(0)}(\omega) t_R {\bf g}_{k,\sigma}(\omega)$, where
$G_\sigma^<(t) = i\langle c^\dagger_\sigma c_\sigma(t)\rangle$ is the Green's
functions of the dot, and
$g_{k,\sigma}^< = i\langle a^\dagger_{k\sigma R} a_{k\sigma R}(t)\rangle$
for the leads, and the retarded and advanced Green's functions are defined
similarly.
We do the analog for $G^{<(0)}_{kd,\sigma}$.
For the (noninteracting) leads we make use of
$g_{k,\sigma}^<(\omega) = 2\pi i f_R(\omega) \delta (\omega - \epsilon_{k})$,
$g_{k,\sigma}^r(\omega) = 1/(\omega - \epsilon_k + i 0^+)$, and
$g_{k,\sigma}^a(\omega) = \left( g_{k,\sigma}^r(\omega) \right)^*$.

We obtain then
\begin{equation}
   I_R^{(0)} = -{ie\over h} \Gamma_R \sum_{\sigma} \int d \omega
        \left[ G_{\sigma}^{< (0)} + f_R \left(
                G_{\sigma}^{r (0)} - G_{\sigma}^{a (0)}
        \right) \right] \, .
\end{equation}
As a check we verify that in equilibrium, $V=0$, the relation
$G^<(\omega) + f(\omega) [G^r(\omega) - G^a(\omega)]=0$ guarantees that no
current is flowing, $I_R=0$.

It is straightforward to derive an analogous expression for $I_L^{(0)}$ for
the left lead.
The total current must be conserved, $I_L^{(0)}+I_R^{(0)}=0$.
Even more, the current must be conserved for each spin separately.
This yields the condition
\begin{eqnarray}
\label{current conservation 0}
   0 = \int d \omega \left[ G_{\sigma}^{< (0)} +
        {\Gamma_L f_L + \Gamma_R f_R \over \Gamma_L + \Gamma_R}
        \left(  G_{\sigma}^{r (0)} - G_{\sigma}^{a (0)}\right) \right]
\end{eqnarray}
for the integral (but not for each energy $\omega$ separately).
Due to current conservation we can write $I_R = (\Gamma_L I_R - \Gamma_R I_L) /
(\Gamma_L + \Gamma_R)$ and find the well-known result for the current
through a QD (in the absence of a reference arm),
\begin{equation}
   I_R^{(0)} = -{2e\over h}{\Gamma_L \Gamma_R\over \Gamma_L+\Gamma_R}
        \sum_{\sigma} \int d \omega \, {\rm Im} \, G_{\sigma}^{r(0)}(f_L - f_R)
        \, ,
\label{current_0}
\end{equation}
and $A_\sigma(\omega) = -(1/\pi)\, {\rm Im} \,G_\sigma^r(\omega)$ is the
spectral density.

We now turn to the flux-dependent term $I_R^{(1)}$.
Similarly to the above, we can make use of Dyson-like equations.
We get a contribution from the first line of Eq.~(\ref{current_start}) and use
${\bf G}^{(0)}_{qk,\sigma} (\omega) = {\bf g}_{q,\sigma}(\omega) t_L
{\bf G}_\sigma^{(0)}(\omega) t_R {\bf g}_{k,\sigma}(\omega)$.
For the contributions of the second line of Eq.~(\ref{current_start}) we make
use of
${\bf G}^{(1)}_{dk,\sigma} (\omega) = {\bf G}_\sigma^{(0)}(\omega) t_L
{\bf g}_{q,\sigma}(\omega) {\tilde t} {\bf g}_{k,\sigma}(\omega)
+ {\bf G}_\sigma^{(1)}(\omega) t_R {\bf g}_{k,\sigma}(\omega)$.
After collecting all terms and employing the relation
$4\pi^2 \tilde t t_R t_L N_L N_R = \sqrt{\Gamma_L \Gamma_R} |t^{\rm ref}|
e^{i\varphi}$ we get\cite{note2}
\begin{eqnarray}
    I_R^{(1)} &=&
        {e \over h} \sqrt{\Gamma_L \Gamma_R} |t^{\rm ref}|
        \sum_{\sigma} \int d \omega
\nonumber \\ &&
        \left[ \cos \varphi
        \left( G_\sigma^{r (0)} + G_\sigma^{a (0)} \right) (f_L - f_R)
\right. \nonumber \\ && \left.
        + i \sin \varphi \, \left( G_\sigma^{<(0)}
        + f_L \left( G_\sigma^{r (0)} - G_\sigma^{a (0)} \right) \right)
        \right]
\nonumber \\ &&-
        {ie \over h} \Gamma_R \sum_{\sigma} \int d \omega \,
        \left[ G_{\sigma}^{< (1)} + f_R \left( G_{\sigma}^{r (1)}
        - G_{\sigma}^{a (1)} \right) \right] \, .
\end{eqnarray}

Again, we can check that due to the equilibrium relation
$G^<(\omega) + f(\omega) [G^r(\omega) - G^a(\omega)]=0$ the current vanishes
at zero bias voltage.
The current conservation (for each spin) in first order in $t^{\rm ref}$ is
equivalent to
\begin{eqnarray}
\label{current conservation 1}
   {\sqrt{\Gamma_L \Gamma_R} \over \Gamma_L+\Gamma_R} |t^{\rm ref}|
        \sin \varphi \int d \omega \,
        \left( G_{\sigma}^{r (0)} - G_{\sigma}^{a (0)} \right) (f_L - f_R)
\nonumber \\
        = \int d \omega \left[ G_{\sigma}^{< (1)} +
        {\Gamma_L f_L + \Gamma_R f_R \over \Gamma_L + \Gamma_R}
        \left(  G_{\sigma}^{r (1)} - G_{\sigma}^{a (1)}\right) \right] \, .
\end{eqnarray}
Note that to ensure current conservation, Green's functions of higher order in
$t^{\rm ref}$ are important since the first line of
Eq.~(\ref{current conservation 1}) alone is in general nonzero (comparison with
Eq.~(\ref{current_0}) shows that it is even proportional to the
current $I^{(0)}$ in zeroth order).
This means that for a systematic and consistent description the influence of
the reference arm on the QD has to be taken into account properly.
It is, in general, not sufficient to treat the Green's function of the QD
independent of the surrounding environment in the device.
One has, therefore, to check carefully in each case up to what extent the
interference signal probes the properties of the QD.

We use $I_R = (\Gamma_L I_R - \Gamma_R I_L) / (\Gamma_L + \Gamma_R)$ and the
current conservation in zeroth order, Eq.~(\ref{current conservation 0}), to
get
\begin{eqnarray}
   I_R^{(1)} &=& {2e\over h} \sqrt{\Gamma_L \Gamma_R} |t^{\rm ref}|
        \cos \varphi \sum_{\sigma} \int d \omega \,
        \, {\rm Re} \, G_{\sigma}^{r (0)} (f_L - f_R)
\nonumber \\ &&-
   {2e \over h} {\Gamma_L \Gamma_R \over \Gamma_R + \Gamma_L} \sum_{\sigma}
        \int d \omega \, {\rm Im}\, G_{\sigma}^{r (1)} (f_L - f_R) \, .
\label{current_1}
\end{eqnarray}
This is the central result of the present derivation.
Again, we emphasize that the influence of the reference arm on the QD plays
a role (it shows up in the Green's functions which are of first order in
$t^{\rm ref}$).
Furthermore, we remark that the Green's functions $G^{(0)}$ and $G^{(1)}$
include multiple tunneling between QD and leads, i.e., they include
contributions from arbitrary high order in $t$ (the superscripts $(0)$ and $(1)$
only labels the order in $t^{\rm ref}$).

\subsection{Linear-response regime}

In the linear-response regime, we replace $f_L(\omega) - f_R(\omega)
\rightarrow -eV f'(\omega)$ in Eq.~(\ref{current_1}) and take the Green's
functions at equilibrium.
The first line in Eq.~(\ref{current_1}) is obviously symmetric under reversal
of magnetic flux, $\varphi \rightarrow - \varphi$.
Moreover, in equilibrium also the Green's functions $G^{(1)}$ have this
property.\cite{note3}
This establishes phase locking in linear response, as expected for a
two-terminal device.

The situation simplifies further if we concentrate on the dominant
contribution to the flux-dependent linear conductance, which is of first order
in $\Gamma$ and first order in $t^{\rm ref}$.
In this case the second line of Eq.~(\ref{current_1}) drops out (since there
is a prefactor of order $\Gamma$ and the equilibrium Green's function
$G^{(1)}$ in zeroth order in $\Gamma$ vanishes).
The connection to the transmission is established by
Eq.~(\ref{nonlinear_current}).
We obtain
\begin{eqnarray}
\label{trans_1a}
   T^{\rm flux}_\sigma(\omega) = 2\sqrt{\Gamma_L \Gamma_R} |t^{\rm ref}|
        \cos \varphi \, {\rm Re} \, G^{r(0)}_\sigma (\omega) \,  ,
\end{eqnarray}
and realize that up to the given order the influence of the reference arm on the
QD does not matter for the transport properties.

For comparison we cite the result for a system with an open
geometry,\cite{Gerland00}
\begin{eqnarray}
\label{trans_1b}
   T^{\rm flux,open}_\sigma(\omega) = 2\sqrt{\Gamma_L \Gamma_R} |t^{\rm ref}|
        \, {\rm Re} \left[ e^{-i \theta} G^{r(0)}_\sigma (\omega) \right]\,  ,
\end{eqnarray}
with $\theta = \varphi +\Delta\theta$, where $\Delta\theta$ is determined by
the specifics of the interferometer.
The two expressions Eqs.~(\ref{trans_1a}) and (\ref{trans_1b}) look very
similar.
One should keep in mind, though, that in higher orders in $\Gamma$ or in
nonlinear response an additional term, which originates from the second line
of Eq.~(\ref{current_1}) enters, while for the derivation of
Eq.~(\ref{trans_1b}) one considers a special case of open geometry where the
reference arm has no effect on the QD's properties.

The transmission Eq.~(\ref{trans_1a}) is always extremal at $\varphi=0$.
Such a ``phase locking'' does not take place in the open-geometry setup:
the AB phase at which the transmission is extremal can be continuously varied
by tuning the energy of the dot level via a gate electrode.

\subsection{Probing the total spin of an interacting QD}

In linear response and lowest order in $t^{\rm ref}$ and $\Gamma$ we get
$G^{r(0)}_\sigma (\omega) = 1 / (\omega -\epsilon + i0^+)$ for $U=0$ and,
according to Eq.~(\ref{trans_1a}), the amplitude of the flux-dependent linear
conductance is symmetric around $\epsilon = 0$.
This is consistent with our intuitive picture, since in the absence of
spin-flip processes no incoherent channels are involved on either side
of the resonance.

This changes when electron-electron interaction in the QD is important.
For $U=\infty$ the retarded Green's function
\begin{equation}
   G^{r(0)}_\sigma (\omega) = {1 \over 1 + f(\epsilon)} \cdot
        {1 \over \omega -\epsilon + i0^+}
\end{equation}
acquires a prefactor $P_0+P_\sigma = 1/[1+f(\epsilon)]$ in linear response and
lowest order in $t^{\rm ref}$ and $\Gamma$.
This prefactor indicates an ``interaction-induced'' asymmetry associated with
spin-flip processes.

\begin{figure}
\centerline{\includegraphics[width=8cm]{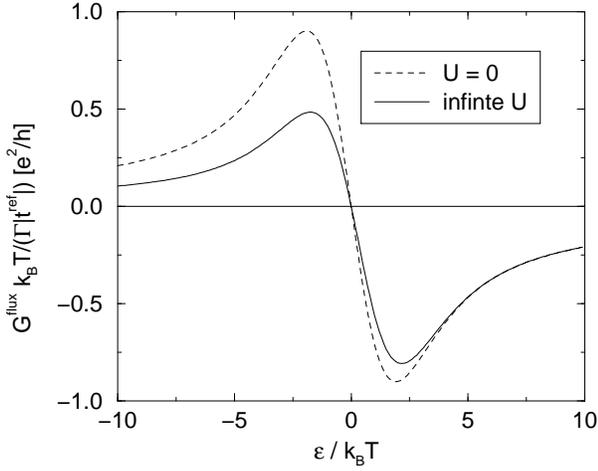}}
\caption{Interference signal for AB interferometer with a single QD.
  Asymmetry in the magnitude of the signal appears for nonzero $U$.
  Plotted is the flux-dependent part of the conductance, normalized by
  $|t^{\rm ref}| \Gamma / (k_BT)$, in units of $e^2/h$.}
\label{fig5}
\end{figure}

For illustration we show in Fig.~\ref{fig5} the flux-dependent linear
conductance as a function of the level position $\epsilon$ for a fixed value
$\varphi = 0$.
In order to obtain the flux dependence, the result simply has to be multiplied
by $\cos \varphi$.
At $\epsilon = 0$ the AB oscillations vanish.
[Note that we consider here only the first harmonic Eq.~(\ref{current_1}).
Higher harmonics survive, an effect known as frequency doubling.]
In the absence of interaction, $U=0$, the amplitude of the AB signal is
symmetric around that point, while for $U=\infty$ clear asymmetry is
predicted.
When the QD is most likely filled up with one electron, the total spin
of the QD is $1/2$ and spin-flip processes reduce the interference signal.
On the other side of the resonance, $\epsilon > 0$, the QD is most likely empty.
The total spin is $0$ and the AB oscillations are by a factor of $2$ larger
than for $\epsilon < 0$.

It is well known that due to the Kondo effect the transport through a QD
increases at low temperature for $\epsilon <0$ but not for $\epsilon > 0$.
Therefore, the Kondo effect reveals the same information as our procedure.
We note, however, that the Kondo effect occurs under much more stringent
experimental conditions, namely, strong coupling of the QD to the leads and
low temperatures (in comparison to the Kondo temperature).
In contrast, our approach to detect the spin of the QD works in the
experimentally easier accessible regime of weak coupling and high temperatures.

\subsection{Coherence of non-spin-flip cotunneling}

As has been outlined in the introduction we use the following procedure to
investigate the coherence of cotunneling.
We tune the system away from resonance, where transport though the QD is
dominated by the usual cotunneling processes.
We then try to find an AB setup such that the total conductance {\it vanishes}
once a reference arm is {\it added}, i.e., we look for complete destructive
interference of the coherent channels.
(The existence of AB oscillations alone only proves {\it partial}
coherence.)
We remind ourselves that the transport through the AB interferometer probes
many channels, characterized by energy $\omega$ and spin $\sigma$,
simultaneously.
Zero transport can only be achieved when {\it all} channels show destructive
interference at the same time.
To achieve this we want to adjust the magnitude of the amplitude $t^{\rm ref}$
for transmission through the reference arm such that
$T^{\rm dot}_\sigma(\omega) = T^{\rm ref}_\sigma$ for {\it all} contributing
energies.

Cotunneling is the dominant transport channel when
$|\epsilon| \gg \Gamma, k_B T$ applies.
In this regime, the linear-response conductance through the QD in the absence of
the reference arm is
\begin{equation}
   {\partial I^{\rm dot} \over \partial V}\bigg|_{V=0} = {2e^2\over h}
        {\Gamma_L \Gamma_R \over \epsilon^2}
\end{equation}
for both $U = 0$ and $U = \infty$.
Cotunneling through the QD is of second order in $\Gamma$.
The conductance through the reference arm in the absence of the QD is
\begin{equation}
   {\partial I^{\rm ref} \over \partial V}\bigg|_{V=0} = {2e^2\over h}
        |t^{\rm ref}|^2 \, .
\end{equation}
We now make the adjustment $|t^{\rm ref}| = \sqrt{\Gamma_L \Gamma_R} /
|\epsilon|$, i.e., $|t^{\rm ref}|$ is of order $\Gamma$.
Therefore, the Green's function which enters the flux-dependent transmission
Eq.~(\ref{trans_1a}) is of zeroth order in $\Gamma$ and $t^{\rm ref}$, i.e.,
the influence of the reference arm on the QD is not probed under the present
conditions.
We find for the total conductance
\begin{equation}
\label{cotunneling noninteracting}
   {\partial I^{\rm tot} \over \partial V}\bigg|_{V=0} = {4e^2\over h}
        {\Gamma_L \Gamma_R \over \epsilon^2} \left[
        1 - {\epsilon \over |\epsilon|} \cos \varphi \right]
\end{equation}
for the noninteracting case and
\begin{equation}
\label{cotunneling interacting}
   {\partial I^{\rm tot} \over \partial V}\bigg|_{V=0} = 4 {e^2\over h}
        {\Gamma_L \Gamma_R \over \epsilon^2} \left[
        1 - {\epsilon \over |\epsilon|} {\cos \varphi \over 1+f(\epsilon)}
        \right]
\end{equation}
for $U=\infty$.
This shows that cotunneling in the noninteracting case is fully coherent.
In the interacting case spin-flip processes are present which spoils coherence.
This is described by the asymmetry factor $1/[1+f(\epsilon)]$, in accordance
with our intuitive picture.

We conclude that an AB interferometer containing a single QD is suitable to prove
coherence of non-spin-flip cotunneling through the QD.

\subsection{Many levels}
\label{subsection_many_levels}

\begin{figure}
\centerline{\includegraphics[width=8cm]{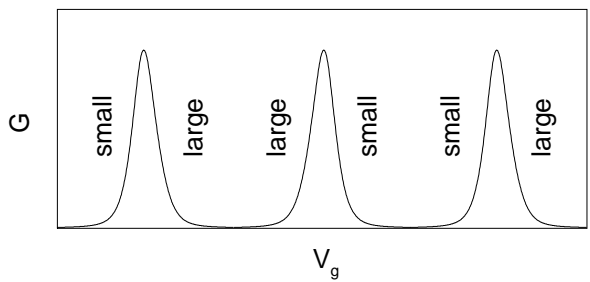}}
\vspace*{.2cm}
\caption{Sequence of asymmetry of AB oscillations in the Coulomb blockade
  regime.
  The solid line depicts schematically the conductance oscillations vs. the gate
  voltage $V_g$, and shows the Coulomb peaks.
  Regions of small and large AB amplitudes are indicated, showing the asymmetry
  between two adjacent Coulomb blockade valleys
  (``large'' corresponds to a valley with a total spin 0, while ``small''
  corresponds to an $S=1/2$ valley).}
\label{fig6}
\end{figure}

In the model we discussed so far, the spacing $\Delta$ of the dot levels was
assumed to be larger than the charging energy and the energy scale provided by
the temperature, the intrinsic linewidth or the voltage bias, so that only one
dot level participates in transport.
We found that the ratio of the number of coherent to the total number of
cotunneling channels was $1$ or $1/2$ in the valley where the electron number
of the dot is even or odd, respectively, giving rise to an observable asymmetry
in the interference signal.
The resulting sequence of asymmetric AB oscillations is shown schematically in
Fig.~\ref{fig6}.
For a continuum of levels, such as in metallic quantum dots, $\Delta=0$,
incoherent cotunneling dominates and we expect little asymmetry.
Dephasing due to electron-electron interactions in the dot is beyond the scope
of the present analysis, but is known\cite{Altshuler97} to be inefficient.

To investigate how the crossover occurs we count the number of coherent and
incoherent cotunneling channels for a system with arbitrary level spacing
$\Delta$.
We examine four different situations.
In all of them the level energy $\epsilon$ is tuned away from resonance,
$k_BT, \Gamma \ll |\epsilon| < U/2$.
The four situations arise due to the combination of being in a valley which
corresponds to an even or odd electron number of the dot and being on either
side of the resonance, i.e., only particle-like or hole-like cotunneling is
considered.

When the level spacing $\Delta$ becomes comparable to $|\epsilon|$ the number
of dot levels involved in transport is increased from $1$ to
$N\sim 2|\epsilon|/\Delta+1$ (each of them is spin degenerate).
As long as $\Delta \gg k_BT$, no inelastic cotunneling process takes place which
changes the energy of the QD.
Incoherence occurs only due to spin flip.
In the ``even'' valley all channels are coherent while in the ``odd'' valley
$1$ out of $N+1$ corresponds to spin flip and is, therefore, incoherent
(the total number of channels is always the same).
The relative asymmetry around the resonance peaks is $N/(N+1)$ and vanishes
for large $N$, i.e., $\Delta \ll |\epsilon|$.

Now we reduce $\Delta$ further such that $\Delta \ll k_BT$.
The number of levels within the range defined by temperature is
$M\sim 2k_B T/\Delta \gg 1$ (each of them is spin degenerate).
For a simple estimate we assume that all single-particle states within the
energy range $\pm k_BT$ around the Fermi energy in the leads are occupied with
probability $1/2$, all states above that range are empty, and all states below
are filled.
For the single-particle states in the QD we do the same.
Let $\delta$ be the level spacing in the leads with $\delta \ll \Delta$.
The number of coherent channels (times the probabilities that the corresponding
states are empty or occupied) is then
\begin{equation}
   {\Delta \over \delta}\cdot {NM \over 2} \, .
\end{equation}

Next, we count the number of incoherent channels where an excitation of energy
$E$ is left on the QD ($E$ is an integer multiple of $\Delta$).
For $E>0$ and given the spin of the incoming and outgoing electron there are
$[M+3 (E/\Delta)]/4$ combinations to leave this excitation on the QD (where the
probabilities to find the corresponding state empty or filled are already
included).
For $E<0$ (i.e., energy is pulled out of the QD) we have $[M+ (E/\Delta)]/4$
combinations.
Energy conservation requires that the energy stored in the QD is pulled out of
the leads and vice versa.
An analogous consideration yields for $E>0$ (energy pulled out of the leads)
$(\Delta/\delta)[M+ (E/\Delta)]/4$ combinations and for $E<0$ we get
$(\Delta/\delta)[M+3 (E/\Delta)]/4$.
In total we find, therefore,
\begin{equation}
   {1\over 2} \cdot {\Delta\over \delta} \cdot \sum_{n=1}^M (M+3n) (M+n)
        = 2 {\Delta\over \delta} \left[ M^3 +
        {\cal O}\left(M^2\right)\right]
\end{equation}
channels with either $E>0$ or $E<0$ (a multiplication factor of 4 was introduced
to account for the spin of the incoming and outgoing electrons).
The number of incoherent channels with $E=0$ (and spin flip) is of the order
$(\Delta/\delta)M^2$ and can be neglected.

The ratio of the number of coherent channels (for $N \gg M \gg 1$) to the
total number of transmission channels is $N/(N+ 4 M^2)$.
As a consequence the coherent contribution vanishes for
$k_BT \gg \sqrt{|\epsilon| \Delta /8}$ or $8(k_BT)^2/|\epsilon| \gg \Delta$.

In summary, by reducing the dot level spacing $\Delta$ we go through three
different regimes:\\
(i) $\Delta \gg |\epsilon|$:
the asymmetry due to spin-flip cotunneling is as large as possible;\\
(ii) $|\epsilon| \gg \Delta \gg 8(k_BT)^2/|\epsilon|$:
the asymmetry vanishes but transport is dominated by coherent channels;\\
(iii) $8(k_BT)^2/|\epsilon| \gg \Delta$:
transport is dominated by incoherent channels.

\subsection{First-order transport}

The discussion so far covered cotunneling, which dominates away from resonance.
Let us now turn to the regime $k_BT \gg \Gamma, |\epsilon|$, where the dot
level is near resonance and transmission is dominated by first-order transport
in $\Gamma$.
Can we tune the single-dot AB interferometer in such a way that full
destructive interference is achieved for all those first-order transport channels
that are coherent?
The energy spread of electrons going through the reference arm is $k_BT$,
while the width of the resonance through the QD is $\Gamma$.
Hence, the temperature has to be on the one hand larger than $\Gamma$ in order
to be in the regime where first-order transport dominates, yet, on the other
hand, it has to be smaller than $\Gamma$ to allow for a destructive
interference of {\em all} energy components simultaneously.
To circumvent this problem, we consider a two-terminal AB interferometer with
two QDs, one in each arm, see Fig.~\ref{fig2}c.

\section{Interferometry with two QDs}
\label{section_two_dots}

The conceptual difficulty to address first-order transport in a single-dot
AB interferometer is that the temperature has to be on the one hand large, yet,
on the other hand, it has to be small to allow for a destructive interference
of all energy components simultaneously.
As we will show in this section, this difficulty will not arise in
AB interferometers which contain {\it two} QDs, one in each arm, see
Fig.~\ref{fig2}c.
This is due to the fact that the resonance of width $\Gamma$ of each QD
filters out a small fraction of the incoming electrons in {\rm both} arms,
even at high temperature, and we will find that fully destructive interference
(in the absence of interaction) is feasible.
For interacting QDs we find again an asymmetry factor which makes it possible
to distinguish on which side of the resonance the adjacent Coulomb blockade
valley has an even or odd number of electrons.

Geometries similar to the two-dot AB interferometer have been investigated
theoretically in the literature.
Resonant tunneling (in the absence of interaction)\cite{Shahbazyan94}
and cotunneling\cite{Akera93,Loss00} have been studied
in this geometry.
Furthermore, a numerical renormalization-group analysis has been
performed\cite{Izumida97} where, by construction, only equilibrium properties
of the dot were included.
Spectral properties of such a double-dot system, or equivalently a two-level
dot, have been addressed.\cite{Boese00,KGS98}
Recently, also experimental realization of a two-dot AB setup has been
reported.\cite{Wiel00,Holleitner00}

\subsection{General current formula}

The total Hamiltonian
\begin{equation}
   H = H^{{\rm dot},1} + H^{{\rm dot},2}
\end{equation}
is the sum of two parts.\cite{note4}
Each of them describe a QD coupled to (the same) reservoirs and has the
structure of Eq.~(\ref{Hamiltonian}).
We choose a completely symmetric geometry, and we assume
$k_B T \gg \Gamma,|\epsilon_1|,|\epsilon_2|$ as well as
$\Gamma \gg |\epsilon_1 -\epsilon_2|$, where $\epsilon_{1,2}$ is the energy of
the level in QD 1 and 2.
In this regime lowest-order transport dominates, and we can set
$\epsilon = \epsilon_1 = \epsilon_2$.
To model the enclosed flux we attach a phase factor $e^{i\varphi/4}$ to the
tunnel matrix elements $t_{R, \rm dot 1}$ and $t_{L, \rm dot 2}$, and
$e^{-i\varphi/4}$ to $t_{L, \rm dot 1}$ and $t_{R, \rm dot 2}$.
This symmetric flux dependence of the tunnel matrix elements can be achieved
choosing a corresponding gauge.

The system is equivalent to a single QD with two levels (each of them spin
degenerate) with $\varphi$-dependent tunnel matrix elements.
To write the total current in a compact way we employ a $2\times 2$ matrix
notation to account for the two QDs and get\cite{Meir92}
\begin{eqnarray}
   I^{\rm tot} = {ie \over 2h} && \int d \omega \, {\rm \bf {tr}} \left\{
        \left[ {\bf \Gamma}^L f_L - {\bf \Gamma}^R f_R \right] {\bf G}^>_\sigma
\right. \nonumber \\ && \left.
        + \left[ {\bf \Gamma}^L (1-f_L) - {\bf \Gamma}^R (1-f_R) \right]
        {\bf G}^<_\sigma
        \right\}
\end{eqnarray}
with ${\bf \Gamma}^{L} = {\Gamma \over 2} \left( \begin{array}{cc}
1 & e^{+ i\varphi/2} \\ e^{- i\varphi/2} & 1 \\ \end{array} \right)
\delta_{\sigma \sigma'}$
and ${\bf \Gamma}^{R} = \left( {\bf \Gamma}^{L}\right)^*$.
The trace has to be performed over both the spin degrees of freedom and the
$2\times 2$ matrices.
We get the general and exact result
\begin{eqnarray}
   I &=& {ie \over 2h} \Gamma \sum_\sigma \int d \omega
\nonumber \\ &&
        \left[
        \left( G^>_{11,\sigma} - G^<_{11,\sigma} +
                G^>_{22,\sigma} - G^<_{22,\sigma} \right)
        \left( f_L - f_R \right)
\right. \nonumber \\ && \left.
        + \cos{\varphi \over 2}
        \left( G^>_{12,\sigma} - G^<_{12,\sigma} +
                G^>_{21,\sigma} - G^<_{21,\sigma} \right)
        \left( f_L - f_R \right)
\right. \nonumber \\ && \left.
        - 2i \sin {\varphi \over 2} \left( {f_L+f_R\over 2} \right)
        \left( G^>_{12,\sigma} - G^>_{21,\sigma} \right)
\right. \nonumber \\ && \left.
        - 2i \sin {\varphi \over 2} \left( 1 - {f_L+f_R\over 2} \right)
        \left( G^<_{12,\sigma} - G^<_{21,\sigma} \right)
        \right]
\label{two dots full}
\end{eqnarray}
with the full Green's functions which includes multiple tunneling through the
entire device.
The indices $1$ and $2$ label the respective QD.
Note that even the diagonal terms $G_{11,\sigma}$ and $G_{22,\sigma}$ are
{\it not} the Green's functions of QD 1 and 2 in the absence of the other one.
Instead, they have to be calculated in the presence of the entire geometry.

\subsection{First-order transport in linear response}

To address the lowest-order contribution to transport we expand
Eq.~(\ref{two dots full}) up to first order in $\Gamma$, i.e., only Green's
functions in zeroth order are involved.
The off-diagonal terms are connected by
$G_{12,\sigma}^{>(0)}(\omega) = G_{12,\sigma}^{<(0)}(\omega) =  2\pi i
P_{2\sigma}^{1\sigma} \delta (\omega-\epsilon)$ to the
stationary off-diagonal density-matrix elements
$P_{2\sigma}^{1\sigma} = \big\langle |2\sigma\rangle \langle 1\sigma|
\big\rangle$ in zeroth order in $\Gamma$.
In equilibrium (and zeroth order in $\Gamma$) the density matrix is diagonal,
with the probabilities determined by Boltzmann weights, and all off-diagonal
matrix elements vanish.
In the first and second line of Eq.~(\ref{two dots full}), only equilibrium
Green's function are involved in linear-response transport.
As a consequence, the second line of Eq.~(\ref{two dots full}) does not
contribute at all (in linear response and first order in $\Gamma$).
In the third and fourth line, however, {\em nonquilibrium Green's functions
enter even in the linear-response regime}.
We obtain
\begin{eqnarray}
   {\partial I^{\rm tot} \over \partial V}\bigg|_{V=0} &=&
        {2 e^2 \over h} \Gamma \sum_\sigma \int d \omega \, \left\{
        {\rm Im} \, G_{11,\sigma}^{r(0)}(\omega) f' (\omega)
\right. \nonumber \\
        &&\left.
        + \sin{\varphi \over 2}
        f(\omega) \, {\partial G^{>(0)}_{12,\sigma}\over \partial (eV)}
\right. \nonumber \\
        &&\left.
        + \sin{\varphi \over 2}
        \left[ 1 - f(\omega) \right]
        {\partial G^{<(0)}_{12,\sigma}\over \partial (eV)}
        \right\} \, .
\label{two dots first order}
\end{eqnarray}
Here, we have used the fact that the contributions involving
${\rm Im} \, G_{22,\sigma}^{r(0)}(\omega)$, $G^{<(0)}_{21,\sigma}(\omega)$,
and $G^{>(0)}_{21,\sigma}(\omega)$ amount to an overall factor 2.
For the first term in Eq.~(\ref{two dots first order}) we use $-(1/\pi)\,
{\rm Im} \, G_{11,\sigma}^{r(0)}(\omega) = \delta (\omega-\epsilon)$ for
$U=0$ and $-(1/\pi)\, {\rm Im} \, G_{11,\sigma}^{r(0)}(\omega) = \delta
(\omega-\epsilon) / [1+f(\epsilon)]$ for $U=\infty$.
This term is flux independent and is twice the conductance through one QD
in the absence of the other one.
Interference effects are accounted for by the second and third terms.
To determine the off-diagonal density-matrix element $P_{2\sigma}^{1\sigma}$
we use the real-time transport theory developed in
Refs.~\onlinecite{Koenig96} and \onlinecite{Koenig99} and solve a generalized 
master equation.
The analysis, presented in Appendix~\ref{append_two_dots}, employs diagrams of
the type used in Appendix~\ref{append_one_dot}.
We find the result (for zeroth order in $\Gamma$ and $V=0$)
\begin{equation}
   {\partial P^{1\sigma}_{2\sigma} \over \partial (eV)}
        = - {i \over 2} f'(\epsilon) \sin (\varphi/2)
\end{equation}
in the absence of interaction and
\begin{equation}
   {\partial P^{1\sigma}_{2\sigma} \over \partial (eV)}
        = - {i\over 2} {f'(\epsilon) \over [1+f(\epsilon)]^3} \sin (\varphi/2)
\end{equation}
for $U=\infty$.
As a consequence, in the absence of an AB flux, only equilibrium Green's
functions enter Eq.~(\ref{two dots first order}).
It is crucial, however, that in the presence of the flux, nonequilibrium
properties of the Green's functions are involved even for the linear-response
regime.

Collecting all terms we find for the noninteracting case
\begin{equation}
\label{two dots noninteracting}
   {\partial I^{\rm tot} \over \partial V}\bigg|_{V=0} =
        2 \, {\partial I^{\rm dot} \over \partial V}\bigg|_{V=0}
        \times \left[ 1 - \sin^2(\varphi/2) \right]
\end{equation}
with $(\partial I^{\rm dot} / \partial V)|_{V=0} = - (\pi e^2 / h)
\Gamma f' (\epsilon)$ being the conductance through a single QD.
At $\varphi= \pm\pi,\pm 3\pi,\ldots$, the total current vanishes (up to a small
correction of order $|\epsilon_1 - \epsilon_2| / \Gamma$, which is not 
considered here),
indicating that lowest-order transport is fully coherent.
In fact it has been shown\cite{Boese00} that in this case, the system can be
mapped onto a model with two levels, where each level is coupled to one of the
leads only.
As a consequence, in the absence of any interaction, there is no way to
transfer any electron from one lead to the other.
This completely contrasts the picture of sequential tunneling which predicts
incoherent transmission.
In the absence of interaction, however, the transport should be fully coherent.
For the simple limit $U=0$, we can rederive Eq.~(\ref{two dots noninteracting})
by using the Landauer-B\"uttiker approach, determining the dot Green's
function by equations of motion, and expanding the result up to first order in
$\Gamma$.
This is done in Appendix~\ref{append_alternative}.
The sequential-tunneling picture is highly misleading in this context.
We should rather view the transport as resonant tunneling but expanded up to
lowest order in $\Gamma$.

In the presence of interaction we obtain
\begin{equation}
\label{two dots interacting}
   {\partial I^{\rm tot} \over \partial V}\bigg|_{V=0} =
        2 \, {\partial I^{\rm dot} \over \partial V}\bigg|_{V=0} \times
        \left[ 1 - {\sin^2(\varphi/2) \over [1+f(\epsilon)]^2} \right]
\end{equation}
with $(\partial I^{\rm dot} / \partial V)|_{V=0} = - (\pi e^2 / h) \Gamma
f' (\epsilon) / [1+f(\epsilon)]$.

We point out that the maximal total conductance (which is reached at
$\varphi =0, \pm 2\pi, \pm 4\pi, \ldots$)
is the sum of the conductances through the QDs taken apart.
There is no extra factor $2$ familiar from constructive interference, which
takes place away from resonance when transmission is small.
To understand this we remark that for $\varphi = 0$ only the symmetric
combination of the dot levels are coupled to the leads while the antisymmetric
one is not (this holds for degenerate levels which are symmetrically coupled
to the leads).
At $U=0$ the model is, therefore, equivalent to a system with one level but
tunnel matrix elements which are $\sqrt{2}$ times the original ones.
As a consequence, in first order in $\Gamma$ (which is second order in the
tunnel matrix elements) the total conductance is twice the conductance
through one arm of the interferometer in the absence of the other one.

The factor $1/[1+f(\epsilon)]^2$ yields an interaction-induced asymmetry in
the ratio of coherent to total transport around a conductance peak.
This factor can be easily understood in the following way.
Consider the situation where the incoming electron has spin up.
Interference is possible if we start with both QDs being empty, one QD
empty and the other one filled with spin up, or both QDs filled with spin up.
For all other starting configurations one would be able to tell afterwards
which way the transfered electron has taken.
The corresponding contributions are not subject to interference.
Summing up the probabilities for the four situations which allow for
interference we find (at $V=0$) the factor $1/[1+f(\epsilon)]^2$.

\begin{figure}
\centerline{\includegraphics[width=8cm]{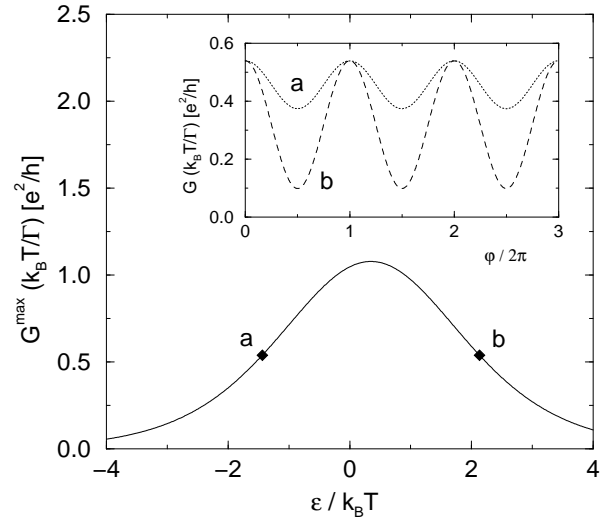}}
\caption{Asymmetry in interference signal for AB interferometer with two QDs.
        Main panel: maximal conductance (i.e., $\varphi = 0$) as a function
        of level position.
        Inset: AB oscillations for the two level positions indicated by a and
        b in the main panel.
        All conductances are normalized by $\Gamma/(k_BT)$ and plotted in units
        of $e^2/h$.}
\label{fig7}
\end{figure}

The effect of the interaction-induced asymmetry factor is illustrated in
Fig.~\ref{fig7}.
In the main panel the maximal first-order linear conductance (which occurs at
$\varphi =0$) is displayed.
In our simple model the resonance peak is symmetric around
$\epsilon /(k_BT) = (1/2) \ln 2$.
Now we pick two level positions at which this maximal conductance is
equal, indicated in the figure by a and b.
The amplitudes of the AB oscillations at these points differ strongly from
each other (inset of Fig.~\ref{fig7}), due to the asymmetry factor
$1/[1+f(\epsilon)]^2$.
The smaller AB amplitude corresponds to the situation in which single
occupation of each QD is more likely than the dots being empty.

\subsection{Effect of dot-dot interaction}

We want to emphasize that for a proper realization of an AB interferometer which
might provide full destructive interference of coherent transport channels,
the two QDs should not influence each other electrostatically.
Let us, e.g., consider the case of spinless electrons in an AB interferometer
in which simultaneous occupation of both QDs costs the
electrostatic-energy penalty $U'$.
Again we start with Eq.~(\ref{two dots first order}), solve for the
corresponding density-matrix elements (see Appendix~\ref{append_two_dots}),
and end up with
\begin{equation}
   {\partial I^{\rm tot} \over \partial V}\bigg|_{V=0} =
        2 \, {\partial I^{\rm dot} \over \partial V}\bigg|_{V=0} \times
        \left[ 1 - C \sin^2(\varphi/ 2) \right]
\end{equation}
with
\begin{equation}
   C = \left[ 1+ {\ln^2 (\beta U'/ 2\pi) \cos^2(\varphi/ 2) \over
        \pi^2 [1-f(\epsilon)]^2} \right]^{-1}
\label{C}
\end{equation}
for $\beta\epsilon \sim 1$ and $\beta U' \gg 1$.
At $\varphi = 0, \pm 2\pi, \pm 4\pi, \ldots$, the total linear conductance is
the sum of the conductances through the dots taken apart.
For $\varphi = \pm \pi, \pm 3\pi, \ldots$, the conductance is zero.
The latter statement is not only true for first-order transport but a general
fact, since it is possible to rotate the basis of the double-dot system such
that one state only couples to the left reservoir and the other one to the
right reservoir.\cite{Boese00}

We observe that now, since the two paths from source to drain are influencing
each other electrostatically, the interference signal no longer provides a
direct tool to distinguish coherent from incoherent transport through a
{\it single} QD.
The amplitude of the oscillations is reduced for increasing $U'$ because an
electron occupying one QD effectively blocks the path through the other QD,
and interference is suppressed.

\section{Finite-bias-induced AB oscillations}
\label{section_finite_bias}

In this section we concentrate on the average number $\langle N \rangle$
of electrons in the QD near resonance, $|\epsilon| \lesssim k_BT$, in the
weak-coupling regime, $\Gamma \ll k_BT$ (and $\Gamma \ll \Delta$).
At equilibrium, $\langle N \rangle$ is, for both the single-dot and two-dot
AB interferometer, determined by classical Boltzmann weights.
The latter depend on the energy of the dot level only.
In particular, they are independent of magnetic flux.
(In the regime specified above, transport is dominated by first order in
$\Gamma$, while the occupation number is described in zeroth order.)
Out of equilibrium, however, a flux dependence of $\langle N \rangle$ might
arise, as we will show in this section.

\subsection{Single-dot AB interferometer}

In the absence of charging energy, the correction terms of the probabilities
for the empty ($\chi=0$), singly-occupied ($\chi=\sigma$ with $\sigma=\uparrow$
or $\downarrow$), and doubly-occupied dot ($\chi=d$), in first order in
$t^{\rm ref}$ are (see Appendix~\ref{append_one_dot})
\begin{equation}
   P_\chi^{(1)} = \alpha_\chi {2\sqrt{\Gamma_L\Gamma_R}\over \Gamma_L+\Gamma_R}
        |t^{\rm ref}| \sin \varphi \left[ f_L(\epsilon) - f_R(\epsilon) \right]
        \, ,
\label{prob}
\end{equation}
with $\alpha_0 = 1 - F (\epsilon)$,
$\alpha_\sigma = F (\epsilon) - 1/2$, and
$\alpha_d = - F (\epsilon)$.
Here we have used the definition
\begin{equation}
   F(\epsilon) = { \Gamma_L f_L(\epsilon) + \Gamma_R f_R(\epsilon) \over
        \Gamma_L+\Gamma_R} \, .
\end{equation}
At finite bias voltage, the probabilities for the dot states depend on the
AB flux {\it even in zeroth order in the intrinsic linewidth $\Gamma$}.
Only at special values of the flux, $\varphi = 0, \pm \pi, \pm 2 \pi, \ldots$,
the correction terms $P_\chi^{(1)}$ vanish for all states
$\chi=0, \uparrow, \downarrow, d$, and the probabilities coincide with those
of a QD without a reference arm.

The same statement is true for interacting QDs.
For $U=\infty$ we obtain correction terms of the type Eq.~(\ref{prob}) with
$\alpha_0 = 1 / [1 + F (\epsilon)]^2$ and
$\alpha_\sigma = - \alpha_0/2$ (see Appendix~\ref{append_one_dot}).

For symmetric couplings, $\Gamma_L = \Gamma_R$, the occupation of the QD up to
first order in $t^{\rm ref}$ is
\begin{equation}
   \langle N \rangle = f_L(\epsilon) + f_R(\epsilon)
        - |t^{\rm ref}| [ f_L(\epsilon) - f_R(\epsilon) ] \sin \varphi
\end{equation}
for noninteracting QDs, $U=0$, and
\begin{equation}
   \langle N \rangle = {f_L(\epsilon) + f_R(\epsilon) \over 1 +
        {f_L(\epsilon) + f_R(\epsilon)\over 2}}
        - {|t^{\rm ref}| [ f_L(\epsilon) - f_R(\epsilon) ] \sin \varphi
        \over \left( 1 + {f_L(\epsilon) + f_R(\epsilon)\over 2} \right)^2 }
\label{induced oscillation one dot}
\end{equation}
for $U=\infty$.

\begin{figure}
\centerline{\includegraphics[width=8cm]{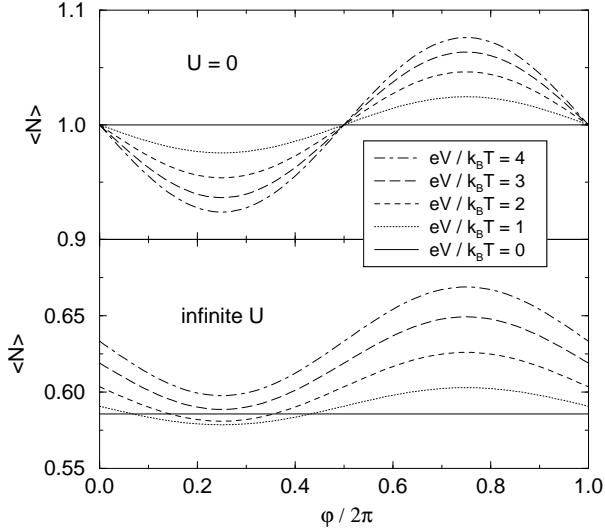}}
\caption{Bias-voltage-induced AB oscillations of the dot occupation in a
        single-dot AB interferometer.
        The level position is tuned at the maximum of the Coulomb-oscillation
        peak.
        In equilibrium the average charge on the QD is flux independent
        (in lowest order in $\Gamma$).
        With finite bias voltage, AB oscillations emerge.
        In this example we chose $|t^{\rm ref}| = 0.1$ and
        $\Gamma_L = \Gamma_R$.}
\label{fig8}
\end{figure}

The generation of AB oscillations due to finite bias is illustrated in
Fig.~\ref{fig8}.
To be specific we chose the level energy $\epsilon$ where the conductance has
its peak, $\epsilon = 0$ for $U=0$ and $\epsilon/(k_BT) = (1/2)\ln 2$.
Furthermore, we used $|t^{\rm ref}| = 0.1$ and chose $\Gamma_L = \Gamma_R$
in this example.

\subsection{Two-dot AB interferometer}

For the AB interferometer containing two QDs it turns out that the corrections
to the diagonal matrix elements of the density matrix in lowest order in
$\Gamma$ remain zero in first order in $V$, i.e., in contrast to the
single-dot AB interferometer, there is no bias-voltage-induced AB oscillation
of the occupation of either dot in linear order in $V$.
This changes when electrostatic inter-dot interaction is introduced (although
it does not change the spatial symmetry).
For the model with spinless electrons and inter-dot interaction we find that
the correction for the occupation of QD 1, $P^1_1$, is determined (for
$\beta \epsilon \sim 1$ and $\beta U' \gg 1$) by
\begin{equation}
        {\partial P^1_1\over \partial V}\bigg|_{V=0}
        = - {C \ln (\beta U' / 2\pi) \over 4\pi [1-f(\epsilon)]^2}\cdot
        {f'(\epsilon) \over 1+f(\epsilon)} \sin \varphi \, ,
\label{induced oscillation two dots}
\end{equation}
where the (also flux-dependent) term $C$ is defined in Eq.~(\ref{C}).
For QD 2 we get
$(\partial P^2_2/ \partial V)|_{V=0} = - (\partial P^1_1/ \partial V)|_{V=0}$.
All these results are derived in Appendix~\ref{append_two_dots}.

\section{Breakdown of phase locking and more general symmetry relations}
\label{section_symmetry}

Onsager relations yield phase locking of the linear conductance through a
(closed) two-terminal device, i.e., the linear conductance is symmetric under
reversal of magnetic flux, $\varphi \rightarrow -\varphi$.
Phase locking is, however, no longer enforced in the nonlinear-response
regime.
In the final result for the interference current through a single-dot
AB interferometer in first order in $|t^{\rm ref}|$, Eq.~(\ref{current_1}),
the first line is symmetric under $\varphi \rightarrow -\varphi$, but the
$\varphi$-dependence of $G^{(1)}$ in the second line can break this symmetry
at finite bias voltages.
In the following we explicitly calculate the current in lowest order in
$\Gamma$.
It will turn out that for a noninteracting QD phase locking survives even at
finite bias, but is broken for interacting QDs.

In the noninteracting case we find to lowest order in $\Gamma$ that
${\rm Im} \, G_{\sigma}^{r (1)} \sim [ P_0^{(1)} + P_\uparrow^{(1)}
+ P_\downarrow^{(1)} + P_{\uparrow\downarrow}^{(1)} ] = 0$, which yields
\begin{equation}
   I_R^{(1)} = {4 e\over h} \sqrt{\Gamma_L \Gamma_R} |t^{\rm ref}|
        \cos \varphi \int ^{\prime} d \omega \,
        {f_L(\omega) - f_R(\omega) \over \omega - \epsilon} \, ,
\end{equation}
where the prime at the integral indicates Cauchy's principal value.
This proves that for a noninteracting QD phase locking is preserved even in
nonequilibrium, at least in lowest order in $\Gamma$.
It can be shown, however, that even in higher orders in $\Gamma$ the Green's
function is invariant under $\varphi \leftrightarrow -\varphi$, i.e., phase
locking is preserved in nonequilibrium.
This is due to the fact that the total current is obtained as a sum over
contributions at different energies, with weights given by the nonequilibrium
conditions.
Each of these contributions satisfies the phase-locking symmetry separately.

This changes when interaction in turned on.
For $U=\infty$ we find that ${\rm Im}\, G_{\sigma}^{r (1)}$ is non zero, and
the current reads
\begin{eqnarray}
   I_R^{(1)} &=& {4 e\over h} {\sqrt{\Gamma_L \Gamma_R} |t^{\rm ref}|
        \cos \varphi \over 1+F(\epsilon)} \int ^{\prime}
        d \omega \, {f_L(\omega) - f_R(\omega) \over \omega - \epsilon}
\nonumber \\
        && + {4\pi e \over h}
        {(\Gamma_L \Gamma_R)^{3/2} |t^{\rm ref}|
        \sin \varphi \over \Gamma^2 [1+F(\epsilon)]^2}
        \left[ f_L(\epsilon) - f_R(\epsilon) \right]^2 \, .
\end{eqnarray}
In addition to the first line, which corresponds to phase locking, there is a
contribution proportional to $\sin \varphi$ (second line) which breaks
the symmetry under $\varphi \leftrightarrow -\varphi$.
As we see it takes both interaction and finite bias voltage to break
phase locking.

For the AB interferometer which involves two QDs, phase locking remains at
finite bias, independent of whether the interaction is included or not.
This is due to the symmetric setup we chose and follows from general
symmetry relations, as discussed next.

\subsection*{Connection between spatial symmetry and symmetry of transport
coefficients}

The general relation for all two-terminal setups
\begin{equation}
   {\partial I (V,\varphi) \over \partial V} =
        {\partial I (-V,-\varphi) \over \partial V} \, ,
\label{time}
\end{equation}
where $V$ is the symmetrically applied bias, yields as a direct consequence the
Onsager relation
\begin{equation}
   {\partial I (\varphi) \over \partial V} \bigg|_{V=0} =
        {\partial I (-\varphi) \over \partial V} \bigg|_{V=0}
\end{equation}
which requires phase locking in linear response.

In addition to this symmetry Eq.~(\ref{time}), spatial symmetries can be
exploited as well.
Let us first consider an AB interferometer with a single QD.
If $\Gamma_L = \Gamma_R$ then the system remains invariant if it is mirrored
with respect to a vertical axis and the chemical potentials of the left and
right leads are exchanged.
This leads only to Eq.~(\ref{time}) and does not induce a new relation.

\begin{figure}
\centerline{\includegraphics[width=8cm]{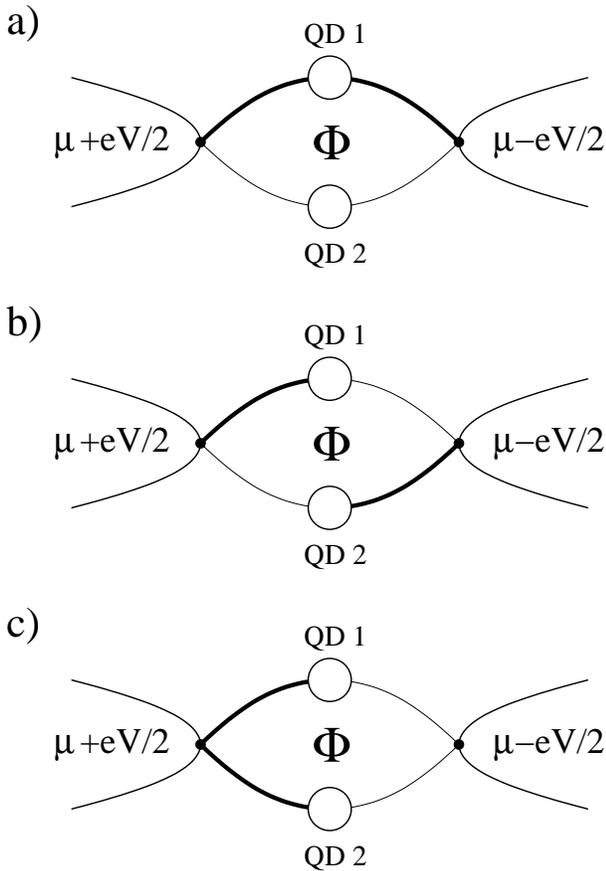}}
\caption{Two-dot AB interferometers with different spatial symmetry.
  Thick and thin lines indicate stronger and weaker tunnel coupling, 
  respectively.
  a) possesses mirror symmetry with respect to a vertical axis, b) is invariant 
  under rotation at angle $\pi$, and c) has mirror symmetry with respect to a 
  horizontal axis.
  In cases b) and c) phase locking is preserved even for finite bias.}
\label{fig9}
\end{figure}

The situation is more complex for the AB interferometer made out of two QDs.
We assume that the QDs are identical.
The coupling of the QDs to the leads, however, may be different.
We consider three different cases in which the system has a spatial symmetry,
see Fig.~\ref{fig9}.
If $|t_{L,{\rm dot} 1}| = |t_{R,{\rm dot} 1}|$ and
$|t_{L,{\rm dot} 2}| = |t_{R,{\rm dot} 2}|$ (see Fig.~\ref{fig9}a) the system has
mirror symmetry with respect to a vertical axis only, and the resulting symmetry
relation
\begin{equation}
   {\partial I (V,\varphi) \over \partial V} =
        {\partial I (-V,-\varphi) \over \partial V}
\label{symmetry 1}
\end{equation}
coincides with Eq~(\ref{time}).

If $|t_{L,{\rm dot} 1}| = |t_{R,{\rm dot} 2}|$ and
$|t_{R,{\rm dot} 1}| = |t_{L,{\rm dot} 2}|$ (see Fig.~\ref{fig9}b)
the system is invariant under rotation at angle $\pi$ in the plane and reversal
of $V$.
As a consequence
\begin{equation}
   {\partial I (V,\varphi) \over \partial V} =
        {\partial I (-V,\varphi) \over \partial V} \, .
\label{symmetry 2}
\end{equation}

If $|t_{L,{\rm dot} 1}| = |t_{L,{\rm dot} 2}|$ and
$|t_{R,{\rm dot} 1}| = |t_{R,{\rm dot} 2}|$ (see Fig.~\ref{fig9}c) the system has
mirror symmetry with respect to a horizontal axis.
This implies
\begin{equation}
   {\partial I (V,\varphi) \over \partial V} =
        {\partial I (V,-\varphi) \over \partial V} \, .
\label{symmetry 3}
\end{equation}
In the two latter cases phase locking occurs, which either follows directly or
after making use of Eq.~(\ref{time}).
This is a consequence of spatial symmetry.
In the first case, or without any spatial symmetry, breaking of phase locking
may be possible for finite voltages.

\section{Summary and Discussion}

The focus of this work is the interplay between coherent and incoherent
transport channels of a mesoscopic setup. 
We specifically address the issue of transport through a quantum dot. 
To stress the difference between dephasing processes (which do not require any 
energy exchange) and inelastic scattering (which, of course, may also introduce 
incoherent transport), we have chosen to discuss spin-flip processes as the 
source of dephasing. 
Our analysis then elucidates some important facets of electronic spin transport.
Central to our analysis is the fact that the existence of decohering spin-flip
channels requires the presence of electron-electron interaction (in our model
the constant capacitance term). 
We have shown explicitly that in the absence of such interaction the 
contributions from the spin-flip channels cancel out.

In order to quantify and study the ``degree of coherence'' in the system we had
to resort to interferometry setups. 
This is why we have embedded our system in an Aharonov-Bohm circuit, in line 
with recent experiments. 
As we present results for the AB amplitude of the transmitted current (as 
compared with the flux-insensitive component of the current), our results are 
closely related the ``visibility'', employed by some experimentalists in 
this context.

Technically  we have performed a systematic analysis of first- and second-order
(in the QD-lead coupling, $\Gamma$)  transport through AB interferometers
containing either one or two QDs.
As a consequence, in the Coulomb blockade valley with one dot level being
singly-occupied only one half of the low-temperature (yet $T>T_K$) transport is 
coherent.
This is to be contrasted with the case of all dot levels being either empty or 
doubly occupied, where transport is fully coherent. 
The conclusions of our analysis, performed for a QD with a single level, are 
generalized to a multilevel QD. 
In contrast to expectations based on a sequential-tunneling picture, we have 
proved that even in first-order transport, the transmission is at least partially
coherent.
Our formalism allows to cover the nonlinear-response regime, too.
We predict bias-voltage-induced AB oscillations in the occupation number of
the QD of a single-dot AB interferometer.

An important outcome of our analysis is the partial suppression of the AB 
conductance oscillations which, in the presence of e-e interactions, is 
asymmetric with respect to the position of the Coulomb peaks. 
This asymmetry is a manifestation of the presence of incoherent transport 
channels, whose number is different between the two sides of a Coulomb peak. 
Our asymmetry effect (calculated in low orders in $\Gamma$) increases with 
lowering the temperature.
The physical picture is expected to change, though, as the temperature is 
lowered towards and beyond the Kondo temperature $T_K$.
In this case, the local spin of the QD is screened by the lead electron spins.
At zero temperature the QD is seen by the transmitted electrons as a pure static 
potential scatterer, i.e., no spin-flip appears.
We therefore expect an increase of the coherent transport contribution when the
temperature is decreased below $T_K$, and a gradual disappearance of our 
asymmetry effect. 
This shows that while our coherence asymmetry effect may provide information 
comparable to that obtained through the Kondo physics (i.e., the QD's spin), the 
latter reflects different physics.
We also note that the asymmetry we refer to concerns the magnitude of the AB 
amplitudes at two points on different sides of the Coulomb blockade peak, which 
correspond to the same total (flux-averaged) conductance.
Our asymmetry {\it does not imply an asymmetric line shape} of the conductance,
and therefore is different from the Fano mechanism.
The presence of such asymmetries in the experiment will be discussed elsewhere.
\cite{note5}
The results of our analysis suggest that one may  use this asymmetry to probe 
the total spin of a QD; we discuss how this asymmetry is suppressed for 
multilevel QDs, as the level spacing is reduced.

We have included in our analysis a brief discussion of time-reversal symmetry
in the problem, specifically the issue of phase locking. 
The latter is usually presented as an outcome of Onsager relations.
We have shown how, in general, it breaks down at finite bias voltages.
In certain cases, however, we find that phase locking is preserved due to
spatial symmetry of the setup.

\section*{Acknowledgement}

We acknowledge helpful discussions with B. Altshuler, N. Andrei, D. Boese, 
P. Coleman, A. de Silva, L. Glazman, Y. Imry, B. Kubala, Y. Nazarov,  Y. Oreg, 
H. Schoeller, and G. Sch\"on.
J.K. acknowledges the Einstein Center for partially supporting his visits to the
Weizmann Institute of Science.
This work was supported by the Deutsche Forschungsgemeinschaft under the
Emmy-Noether program, the U.S.-Israel Binational Science Foundation, the Minerva
Foundation, the Israel Science Foundation of the Israel Academy of Sciences and
Humanities Centers of Excellence Program and by the Deutsch-Israelisches Projekt.

\begin{appendix}

\section{Density matrix for single-dot AB interferometer}
\label{append_one_dot}

In this appendix we present the somewhat technical part of calculating the
density-matrix elements for the single-dot AB interferometer.
The derivation is based on the real-time transport theory developed in
Refs.~\onlinecite{Koenig96,Koenig99}.
The starting point is the generalized stationary master equation in Liouville
space,
\begin{equation}
\label{master general}
   \left( \epsilon_{\chi_1} - \epsilon_{\chi_2} \right) P^{\chi_1}_{\chi_2}
        + \sum_{\chi_1',\chi_2'} P^{\chi_1'}_{\chi_2'}
        \Sigma^{\chi_1',\chi_1}_{\chi_2',\chi_2} = 0 \, ,
\end{equation}
where $\chi_1$ and $\chi_2$ denote any state for the QD, $\epsilon_{\chi_1}$
and $\epsilon_{\chi_2}$ are the corresponding energies, and
$P^{\chi_1}_{\chi_2} = \langle |\chi_2 \rangle \langle \chi_1 | \rangle$ is a
matrix element of the reduced density matrix for the QD subsystem.
The diagonal matrix elements $P_{\chi_1} \equiv P^{\chi_1}_{\chi_1}$ are
nothing but the probabilities to find the system in a given state $\chi_1$.

The matrix elements $P$ of the density matrix are connected to each other
in Eq.~(\ref{master general}) by the terms
$\Sigma^{\chi_1',\chi_1}_{\chi_2',\chi_2}$ which can be viewed as generalized
transition rates in Liouville space.
They are defined as the irreducible self-energy parts of the propagation in
Liouville space and are represented as diagram blocks on a Keldysh contour.
For a detailed derivation of this diagrammatic language, the generalized master
equation, and the rules on how to calculate the value of a diagram we refer to
Refs.~\onlinecite{Koenig96,Koenig99}.

In the present context we calculate the corrections linear in $t^{\rm ref}$ to 
the probabilities for the dot states.
We concentrate on the regime where the dot level is tuned close to resonance,
i.e., we determine the probabilities in zeroth order in $\Gamma$.
To this end we need the self-energy parts $\Sigma$ to lowest (first) order in 
$\Gamma$.
The possible dot states $\chi$ are labeled by $0$ for an empty QD, $\sigma$
for single occupancy with spin $\sigma = \uparrow, \downarrow$, and $d$ for
double occupancy.

\begin{figure}
\includegraphics[width=8cm]{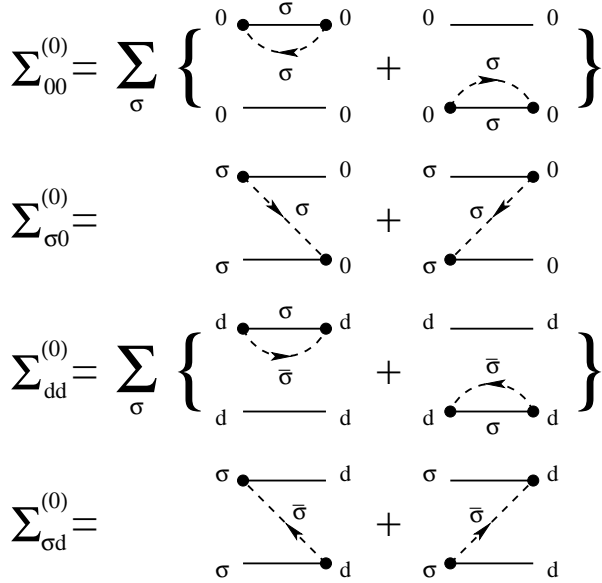}
\caption{Irreducible self-energy parts to zeroth order in $t^{\rm ref}$ (and
  first order in $\Gamma$).}
\label{fig10}
\end{figure}

First, we observe that in the generalized master equation
Eq.~(\ref{master general}) the probabilities $P_\chi = P_\chi^\chi$ couple to
diagonal matrix elements only.
This has to do with the fact that the $z$ component of the spin is a conserved
quantity.
As a consequence, only self-energy parts of the type
$\Sigma_{\chi,\chi'}\equiv \Sigma_{\chi,\chi'}^{\chi,\chi'}$ enter.
We get to zeroth order (indicated by the superscript $(0)$) in $t^{\rm ref}$ the 
master equations
\begin{eqnarray}
  P_0^{(0)} \Sigma_{00}^{(0)} + 2 P_\sigma^{(0)} \Sigma_{\sigma 0}^{(0)} 
  + P_d^{(0)} \Sigma_{d0}^{(0)} &=& 0
\\
  P_0^{(0)} \Sigma_{0d}^{(0)} +
  2 P_\sigma^{(0)} \Sigma_{\sigma d}^{(0)} + P_d^{(0)} \Sigma_{dd}^{(0)} &=& 0
\end{eqnarray}
together with the normalization condition
$P_0^{(0)} + 2 P_\sigma^{(0)} + P_d^{(0)} = 1$.
The diagrammatic representation of the irreducible self-energy parts are shown in
Fig.~\ref{fig10}.
There are two horizontal lines representing the forward (upper line) and backward
(lower line) propagators on the Keldysh contour.
Vertices (full dots) indicate tunneling from the QD to the lead or vice versa.
They are connected in pairs by tunnel lines (dashed lines) representing
contractions of lead electron operators.
Depending on the order of the annihilation and creation operators, specified by
the arrows, the tunnel lines contribute with a Fermi function $f_r$ (if the line
goes backward with respect to the Keldysh contour) or $1-f_r$ (if the line goes 
forward), $r=L/R$, times $|t_r|^2 N_r = \Gamma_r/(2\pi)$, where $N_r$ is the
density of states in reservoir $r$.
Along the tunnel lines and the forward and backward propagators we assign the 
proper dot states such that the particle number and $z$ component of the spin 
are conserved at each vertex.

It is easy to see that it is impossible to construct a diagram for 
$\Sigma_{d0}$ or $\Sigma_{0d}$ to first order in $\Gamma$, i.e., 
$\Sigma_{d0} = \Sigma_{0d} = 0$.
The other diagrams are depicted in Fig.~\ref{fig10}.
As an example we calculate $\Sigma_{\sigma 0}^{(0)}$ explicitly, using the 
diagrammatic rules derived in Refs.~\onlinecite{Koenig96,Koenig99}.
The tunnel line contributes with $\Gamma_r/(2\pi) [1-f_r(\omega)]$.
There is a minus sign for each vertex on the backward propagator.
This yields a total minus sign for the diagram under consideration.
For each segment between two vertices there is a resolvent 
$1 / (\Delta E + i0^+)$, where $\Delta E$ is the difference between the energy on
the left-going lines and right-going lines (including the propagators and the 
tunnel lines).
For the left diagram we obtain $1/(\epsilon - \omega + i0^+)$.
Finally, we integrate over the energy $\omega$ of the tunnel line and sum over
the reservoir index $r$ to get
\begin{equation}
   -\sum_r {\Gamma_r \over 2\pi} \int d\omega \,
   {1-f_r(\omega) \over \epsilon - \omega + i0^+} \, .
\end{equation}
The value of the right diagram is the same except for the denominator which now
reads $\omega - \epsilon + i0^+$.
After making use of $1/(x+i0^+) = P(1/x) - i \pi \delta(x)$ we can evaluate the 
integrals.
The final result (for all diagrams shown in Fig.~\ref{fig10}) is
\begin{eqnarray}
   \Sigma_{00}^{(0)} &=& -2 i \Gamma F(\epsilon)
\\
   \Sigma_{\sigma 0}^{(0)} &=& i \Gamma [1-F(\epsilon)]
\\
   \Sigma_{dd}^{(0)} &=& -2 i \Gamma [1-F(\epsilon+U)]
\\
   \Sigma_{\sigma d}^{(0)} &=& i \Gamma F(\epsilon+U) \, ,
\end{eqnarray}
where we have used the definitions
$F(\omega) = [\Gamma_L f_L(\omega) + \Gamma_R f_R(\omega)] /
(\Gamma_L+\Gamma_R)$ and $\Gamma = \Gamma_L + \Gamma_R$.
This yields the solution
\begin{eqnarray}
   P_0^{(0)} &=& { [1- F(\epsilon)] [1-F(\epsilon+U)] \over
        F(\epsilon) + 1 -F(\epsilon+U) }
\\
   P_\sigma^{(0)} &=& { F(\epsilon) [1-F(\epsilon+U)] \over
        F(\epsilon) + 1 -F(\epsilon+U) }
\\
   P_d^{(0)} &=& { F(\epsilon) F(\epsilon+U) \over
        F(\epsilon) + 1 -F(\epsilon+U) } \, .
\end{eqnarray}
In equilibrium $F(\omega) = f(\omega)$, and we recover the classical Boltzmann
factors which determine the probabilities, $P_0^{(0)} = 1/Z$,
$P_\sigma^{(0)} = \exp(-\beta \epsilon)/Z$, and
$P_d^{(0)} = \exp[-\beta (2\epsilon+U)]/Z$ with
$Z = 1+2\exp(-\beta \epsilon)+\exp[-\beta (2\epsilon+U)]$.

\begin{figure}
\noindent
\includegraphics[width=17.3cm]{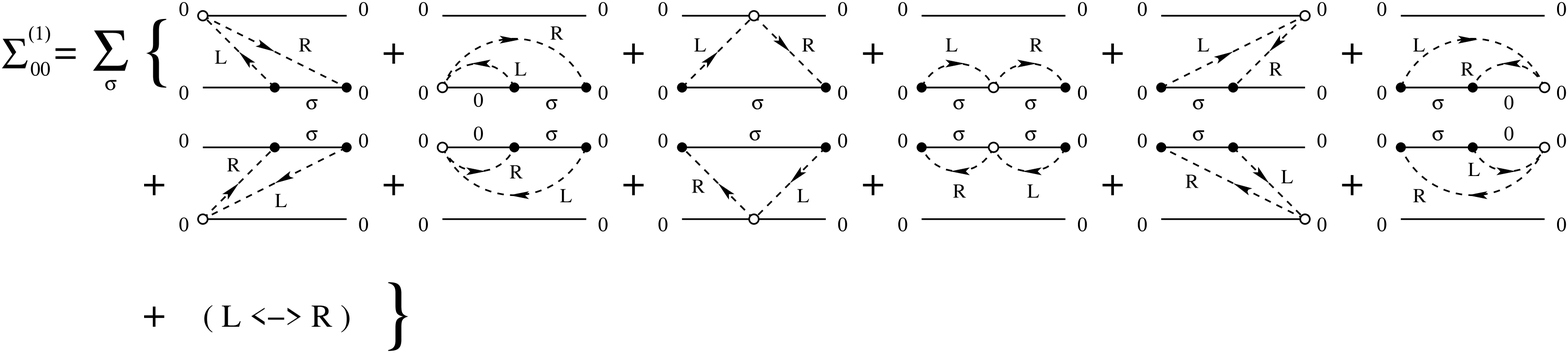}
\includegraphics[width=17.3cm]{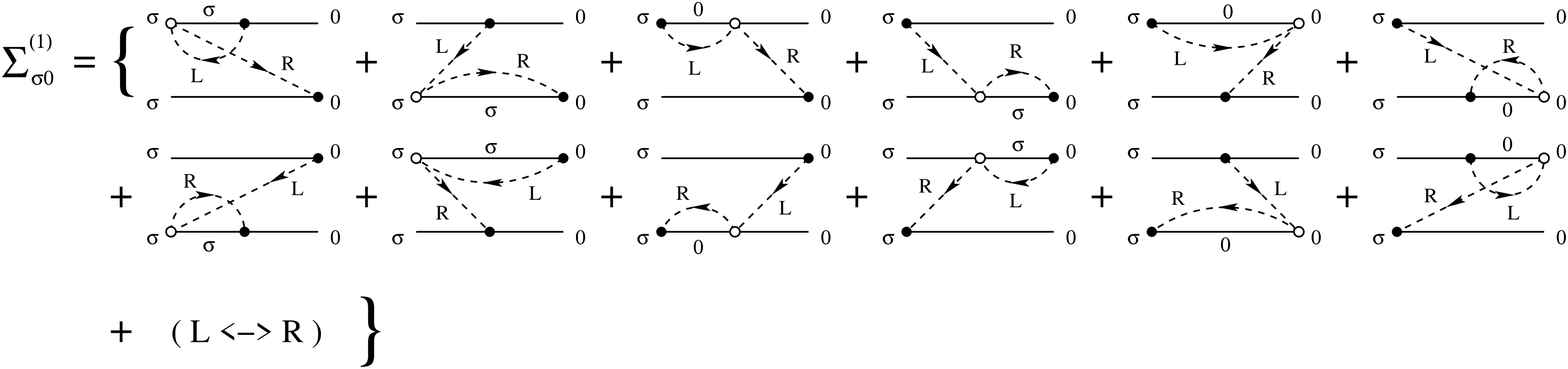}
\includegraphics[width=17.3cm]{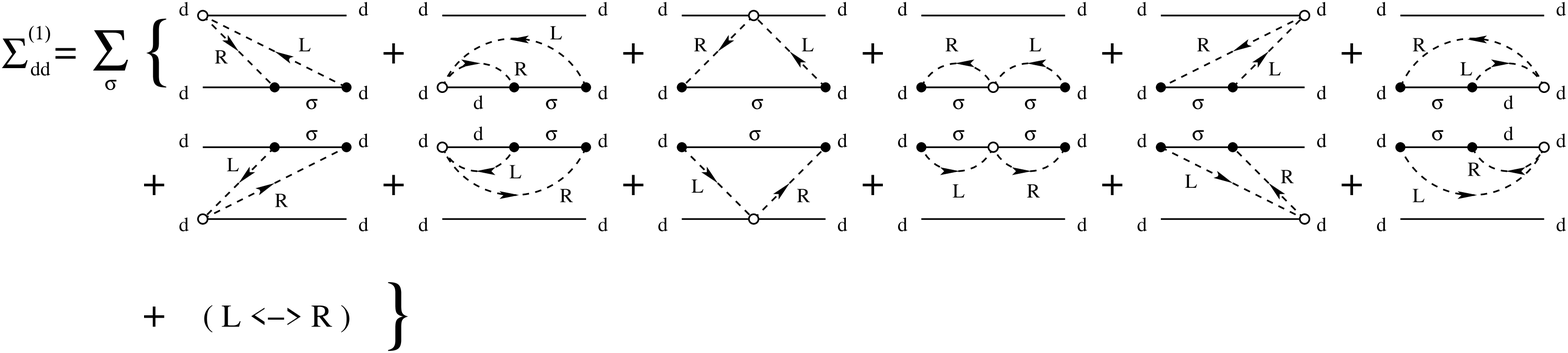}
\includegraphics[width=17.3cm]{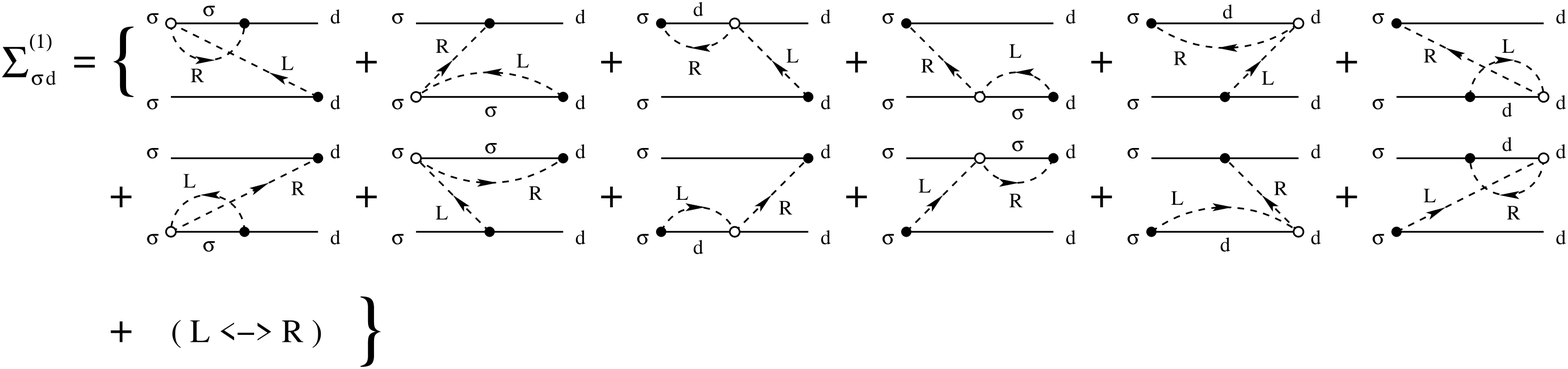}
\caption{Irreducible self-energy parts to first order in $t^{\rm ref}$ (and
  first order in $\Gamma$).
  There are two kinds of vertices: single vertices (full dots), which represent 
  tunneling between QD and leads, and double vertices (open dots), which 
  indicate direct tunneling between the leads.
  We show explicitly all diagrams with direct tunneling from left to right
  (they are proportional to $e^{i\varphi}$).
  The remaining diagrams are obtained by exchanging the labels $L$ and $R$ for
  the left and right lead everywhere (those are proportional to 
  $e^{-i\varphi}$).}
\label{fig11}
\end{figure}

The effect of the reference arm on the QD shows up in the correction terms to
the probabilities in first order in $t^{\rm ref}$.
The master equations to solve are
\begin{eqnarray}
   P_0^{(0)} \Sigma_{00}^{(1)} +
   2 P_\sigma^{(0)} \Sigma_{\sigma 0}^{(1)} +
   P_0^{(1)} \Sigma_{00}^{(0)} +
   2 P_\sigma^{(1)} \Sigma_{\sigma 0}^{(0)} &=& 0
\\
   P_d^{(0)} \Sigma_{dd}^{(1)} +
   2 P_\sigma^{(0)} \Sigma_{\sigma d}^{(1)} +
   P_d^{(1)} \Sigma_{dd}^{(0)} +
   2 P_\sigma^{(1)} \Sigma_{\sigma d}^{(0)} &=& 0
\end{eqnarray}
along with the normalization condition
$P_0^{(1)} + 2 P_\sigma^{(1)} + P_d^{(1)} = 0$.
The diagrams for the self-energy part to first order in $t^{\rm ref}$ are shown
in Fig.~\ref{fig11}.

\pagebreak
\phantom{x}
\pagebreak

Here, another type of vertex has to be introduced which represents direct 
tunneling from one lead to the other.
Such a vertex is connected to two tunnel lines (``double vertex'') since it
represents two lead electron operators which are contracted with another 
operator of the same lead each.
As a consequence we encounter objects made of two tunnel lines connecting one
double vertex with two single vertices.
Each line contributes with $f_r$ or $1-f_r$, and the total multiplication factor
is $N_L N_R t_Lt_R |\tilde t| = |t^{\rm ref}| \sqrt{\Gamma_L \Gamma_R}/(4\pi^2)$ 
times $e^{\pm i\varphi}$.
The upper (lower) sign applies if the incoming line into the double vertex 
belongs to the left (right) lead.

Again we evaluate one diagram explicitly, namely the first one contributing to
$\Sigma_{\sigma 0}^{(1)}$ as shown in Fig.~\ref{fig11}.
We find
\begin{equation}
   \int d\omega_L  \int d\omega_R \,
   {f_L(\omega_L) [ 1 - f_R (\omega_R) ] \over
     (\omega_L - \omega_R + i0^+) (\epsilon - \omega_R + i0^+)}
\end{equation}
times $|t^{\rm ref}| e^{i\varphi} \sqrt{\Gamma_L \Gamma_R}/(4\pi^2)$.
There is one minus sign due to the vertex on the lower propagator.
This is, however, canceled by another minus sign coming from the ordering of
the Fermi operators: in the diagrammatic language, each crossing of tunnel lines 
introduces a minus sign (to apply this rule view the double vertex as two 
separate vertices lying close together, where the vertex connected to the 
incoming line comes first with respect to the Keldysh contour).\cite{note6}

We note that for all self-energy parts shown in Fig.~\ref{fig11} the fifth and 
sixth diagrams cancel each other out since they differ by a minus sign due to 
the vertices on the lower propagator.
The same holds for the eleventh and twelfth diagrams. 

After collecting all contributions we end up with
\begin{eqnarray}
   \Sigma_{\sigma 0}^{(1)} &=&
        i \sin \varphi \sqrt{\Gamma_L \Gamma_R} |t^{\rm ref}|
        \left[ f_L(\epsilon)-f_R(\epsilon)\right]
\\
   \Sigma_{\sigma d}^{(1)} &=&
        - i \sin \varphi \sqrt{\Gamma_L \Gamma_R} |t^{\rm ref}|
        \left[ f_L(\epsilon+U)-f_R(\epsilon+U)\right]
\end{eqnarray}
as well as $\Sigma_{00}^{(1)} = 2 \Sigma_{\sigma 0}^{(1)}$ and
$\Sigma_{dd}^{(1)} =  2 \Sigma_{\sigma d}^{(1)}$.

The results for $P_\chi^{(1)}$ are lengthy expressions.
They simplify, however, for either $U=0$ or $U=\infty$.
For these limits they are given by Eq.~(\ref{prob}).

\section{Density matrix for two-dot AB interferometer}
\label{append_two_dots}

In this appendix we determine the off-diagonal matrix element 
$P_{2\sigma}^{1\sigma}$ in zeroth order in $\Gamma$ for the two-dot AB 
interferometer.
These results are needed for the evaluation of Eq.~(\ref{two dots first order}).
To achieve this goal we expand Eq.~(\ref{master general}) up to zeroth and first 
order in $\Gamma$.
The irreducible self-energy parts $\Sigma$ have contributions of order $\Gamma$
and higher.
As a consequence, the zeroth order expansion of Eq.~(\ref{master general})
yields that all off-diagonal matrix elements $P^{\chi_1}_{\chi_2}$ with
$\chi_1 \neq \chi_2$ can only arise for $\epsilon_{\chi_1}= \epsilon_{\chi_2}$.
Otherwise, $P^{\chi_1}_{\chi_2}$ vanishes.

The calculation in this section is based on the same diagrammatic language as
introduced in the previous appendix. 
In the two-dot AB interferometer there is no direct tunneling between leads, 
i.e., no double vertices will appear in the diagrams.
Since constructing and evaluating the diagrams is straightforward along the line 
indicated for the single-dot AB interferometer in the previous appendix,
we refrain from drawing them explicitly.

\subsection{No interaction}

In the case of noninteracting QDs, $U=0$, the two channels provided by the
spin degree of freedom do not influence each other.
The Hamiltonian is just the sum of two identical models, one for each spin.
In this case it is sufficient to consider a simpler model with spinless
electrons and multiply the final expressions for the current by a trivial factor
of $2$.
This simplifies both the calculations and the notations.

The Hilbert space of the double-dot system is, then, four dimensional:
the double dot can be empty ($\chi=0$), singly occupied with the electron in
dot 1 or 2 ($\chi=1,2$), or both dots are filled ($\chi=12$).
The corresponding energies are $0$, $\epsilon$, and $2\epsilon$, respectively.
The density matrix has 16 matrix elements.
Four of them are the diagonal matrix elements $P_0^0$, $P_{1}^{1}$,
$P_{2}^{2}$, and $P_{12}^{12}$.
They are all real and positive.
Since the transformation $1\leftrightarrow 2$ and $V\leftrightarrow -V$ does
not change the system, the three combinations
$P_0^0$, $P_{1}^{1} + P_{2}^{2}$, and $P_{12}^{12}$ are even in $V$, while the
combination $P_{1}^{1} - P_{2}^{2}$ is odd.
There are two nonvanishing off-diagonal matrix elements, $P_{1}^{2}$ and
$P_{2}^{1}$.
All others are zero in zeroth order of $\Gamma$ since the energies of the
corresponding states differ from each other.
The hermiticity of the density matrix implies
$P_{1}^{2} = \left( P_{2}^{1} \right)^*$.

In equilibrium, $V=0$, the diagonal matrix elements $P_{\chi}^{\chi}$ are
probabilities for the state $\chi$ determined by classical Boltzmann factors
$\exp( -\beta \epsilon_\chi)$, and all off-diagonal matrix elements vanish.
Our goal is to determine the linear corrections in $V$.
Only the elements which are odd in $V$ have a finite correction.
These elements are $P_{1}^{1} - P_{2}^{2}$ and ${\rm Im}\, P_{2}^{1}$.

In the following, we write $P= \bar P + \hat P +\ldots$ and
$\Sigma = \bar \Sigma + \hat \Sigma + \ldots$ for the zeroth and first order in
$V$.
It turns out that there are two independent equations which relate the
first-order corrections $\hat P_{1}^{1} = - \hat P_{2}^{2}$ and
${\rm Im} \, \hat P_{2}^{1} = - {\rm Im} \, \hat P_{1}^{2} $ to the
zeroth-order terms $\bar P_0^0$, $\bar P_{1}^{1}=\bar P_{2}^{2}$, and
$\bar P_{12}^{12}$.

We will make use of the relations
$\bar P_\chi^{\chi'} = \bar P_{\tilde \chi}^{\tilde \chi'}$ and
$\bar \Sigma_{\chi,\chi''}^{\chi',\chi'''} =
\bar \Sigma_{\tilde \chi, \tilde \chi''}^{\tilde \chi',\tilde \chi'''}$
in equilibrium and
$\hat P_\chi^{\chi'} = - \hat P_{\tilde \chi}^{\tilde \chi'}$ and
$\hat \Sigma_{\chi,\chi''}^{\chi',\chi'''} = -
\hat \Sigma_{\tilde \chi, \tilde \chi''}^{\tilde \chi',\tilde \chi'''}$ for the
first-order correction in $V$, where $\tilde \chi$ is obtained from $\chi$ by
the transformation $1\leftrightarrow 2$.
For transition from diagonal states in Liouville space to diagonal ones we find
$\hat \Sigma_{\chi,\chi'}^{\chi,\chi'}=0$.
Finally, we drop all $\Sigma$ terms which connect states in Liouville space that
are not compatible, at least to lowest order in $\Gamma$.
As a consequence the master equations for the linear correction in $V$ read
\begin{eqnarray}
\label{master 1}
 0 &=& \hat P_{1}^{1} \bar \Sigma_{1,1}^{1,1}
        + \hat P_{2}^{1} \left(
        \bar \Sigma_{2,1}^{1,1} - \bar \Sigma_{1,1}^{2,1} \right)
\\
 0 &=& \hat P_{2}^{1} \bar \Sigma_{2,2}^{1,1}
        + \hat P_{1}^{1} \left(
        \bar \Sigma_{1,2}^{1,1} - \bar \Sigma_{2,2}^{2,1} \right)
\nonumber \\
        &&+ \bar P_0^0 \hat \Sigma_{0,2}^{0,1}
        + \bar P_{1}^{1} \left(
        \hat \Sigma_{1,2}^{1,1} + \hat \Sigma_{2,2}^{2,1} \right)
        + \bar P_{12}^{12} \hat \Sigma_{12,2}^{12,1} \, .
\label{master 2}
\end{eqnarray}

We calculate all diagrams explicitly and find
\begin{eqnarray}
   \bar \Sigma_{1,1}^{1,1} &=& \bar \Sigma_{2,2}^{1,1} = - i \Gamma
\\
   \bar \Sigma_{2,1}^{1,1} - \bar \Sigma_{1,1}^{2,1} &=&
   \bar \Sigma_{1,2}^{1,1} - \bar \Sigma_{2,2}^{2,1} = 0
\end{eqnarray}
for $V=0$, and
\begin{equation}
   \hat \Sigma_{0,2}^{0,1} =
        {\hat \Sigma_{1,2}^{1,1} + \hat \Sigma_{2,2}^{2,1} \over 2} =
        \hat \Sigma_{12,2}^{12,1} =
        {\Gamma\over 2} eV f'(\epsilon) \sin {\varphi\over 2}
\end{equation}
for the first-order correction in $V$.
This leads to
\begin{eqnarray}
 \hat P_{1}^{1} &=& 0
\\
 \hat P_{2}^{1} &=& - {i\over 2} eV f'(\epsilon) \sin {\varphi \over 2} \, .
\end{eqnarray}

\subsection{Infinite charging energy}

It is straightforward to generalize the previous discussion to the case of
QDs with infinite charging energy, $U=\infty$.
In addition to the label for dots $1$ and $2$, we use $\sigma$ to label the
spin, and $\bar \sigma$ for the spin opposite to $\sigma$.
Since infinite charging energy suppresses all states where either one or both
dots are occupied with two electrons, the Hilbert space has nine dimensions.
Spin symmetry reduces the number of independent diagonal matrix elements from
nine to five,
$P_0^0$,
$P_{1\sigma}^{1\sigma} = P_{1\bar \sigma}^{1\bar \sigma}$,
$P_{2\sigma}^{2\sigma} = P_{2\bar \sigma}^{2\bar \sigma}$,
$P_{1\sigma 2\sigma}^{1\sigma 2\sigma} =
P_{1\bar \sigma 2\bar \sigma}^{1\bar \sigma 2\bar \sigma}$, and
$P_{1\sigma 2\bar \sigma}^{1\sigma 2\bar \sigma} =
P_{1\bar \sigma 2\sigma}^{1\bar \sigma 2\sigma}$.
They are all real.
Since the transformation $1\leftrightarrow 2$ and $V\leftrightarrow -V$ does
not change the system, the four combinations $P_0^0$, $P_{1\sigma}^{1\sigma} +
P_{2\sigma}^{2\sigma}$,
$P_{1\sigma 2\sigma}^{1\sigma 2\sigma}$, and
$P_{1\sigma 2\bar \sigma}^{1\sigma 2\bar \sigma}$ are even in $V$, while the
combination $P_{1\sigma}^{1\sigma} - P_{2\sigma}^{2\sigma}$ is odd.

We are only interested in those off-diagonal matrix elements for which the
projection on the $z$ direction of the total spin of the corresponding states
is the same.
There are six of them, but again spin symmetry reduced the number of
independent elements to three,
$P_{2\sigma}^{1\sigma} = P_{2\bar \sigma}^{1\bar \sigma}$,
$P_{1\sigma}^{2\sigma} = P_{1\bar \sigma}^{2\bar \sigma}$, and
$P_{1\sigma 2\bar \sigma}^{1\bar \sigma 2\sigma} =
P_{1\bar \sigma 2\sigma}^{1\sigma 2\bar \sigma}$.
Furthermore, $P_{1\sigma}^{2\sigma} = \left( P_{2\sigma}^{1\sigma} \right)^*$.
There are only two independent variables,
$P_{1\sigma}^{1\sigma} - P_{2\sigma}^{2\sigma}$ and
${\rm Im}\, P_{2\sigma}^{1\sigma}$, which have a linear correction in $V$.

Again we write $P= \bar P + \hat P +\ldots$ and
$\Sigma = \bar \Sigma + \hat \Sigma + \ldots$ for the term in zeroth order in
$V$ and the first-order correction.
There are two independent equations which look identical to
Eqs.~(\ref{master 1}) and (\ref{master 2}) if we replace
$1 \rightarrow 1\sigma$, $2 \rightarrow 2\sigma$, and
$12 \rightarrow 1\sigma2\sigma$, i.e., the equations for different spins
decouple.

We calculate all diagrams explicitly and find
\begin{eqnarray}
   \bar \Sigma_{1\sigma,1\sigma}^{1\sigma,1\sigma} &=&
   \bar \Sigma_{2\sigma,2\sigma}^{1\sigma,1\sigma} =
        - i \Gamma \left[ 1 + f(\epsilon) \right]
\\
   \bar \Sigma_{2\sigma,1\sigma}^{1\sigma,1\sigma} -
   \bar \Sigma_{1\sigma,1\sigma}^{2\sigma,1\sigma} &=&
   \bar \Sigma_{1\sigma,2\sigma}^{1\sigma,1\sigma} -
   \bar \Sigma_{2\sigma,2\sigma}^{2\sigma,1\sigma} = 0
\end{eqnarray}
for the terms at equilibrium, and
\begin{equation}
   \hat \Sigma_{0,2\sigma}^{0,1\sigma} =
        { \hat \Sigma_{1\sigma,2\sigma}^{1\sigma,1\sigma} +
        \hat \Sigma_{2\sigma,2\sigma}^{2\sigma,1\sigma} \over 2} =
        \hat \Sigma_{1\sigma 2\sigma,2\sigma}^{1\sigma 2\sigma,1\sigma} =
        {\Gamma\over 2} eV f'(\epsilon) \sin {\varphi \over 2}
\end{equation}
for the corrections in first order in $V$.
This leads to
\begin{eqnarray}
   \hat P_{1\sigma}^{1\sigma} &=& 0
\\
   \hat P_{2\sigma}^{1\sigma}
        &=& -{i\over 2} eV {f'(\epsilon) \over [1+f(\epsilon)]^3}
        \sin (\varphi/2) \, .
\end{eqnarray}

\subsection{Dot-dot interaction, no spin}

Finally we consider the case of a finite interaction between dot 1 and 2.
For this case, again Eqs.~(\ref{master 1}) and (\ref{master 2}) hold.
In the limit $\beta \epsilon \sim 1$ but $\beta U' \gg 1$ we find
\begin{eqnarray}
   \bar \Sigma_{1,1}^{1,1} &=& \bar \Sigma_{2,2}^{1,1} = - i \Gamma \left[
        1 - f(\epsilon) \right]
\\
   \bar \Sigma_{2,1}^{1,1} - \bar \Sigma_{1,1}^{2,1} &=&
   \bar \Sigma_{1,2}^{1,1} - \bar \Sigma_{2,2}^{2,1} = {\Gamma\over \pi}
        \ln {\beta U'\over 2\pi} \cos{\varphi\over 2}
\end{eqnarray}
in equilibrium, and
\begin{eqnarray}
   \hat \Sigma_{0,2}^{0,1} &=&
        \hat \Sigma_{1,2}^{1,1} + \hat \Sigma_{2,2}^{2,1} =
        {\Gamma \over 2} eV f'(\epsilon) \sin {\varphi\over 2}
\\
        \hat \Sigma_{12,2}^{12,1} &=& 0
\end{eqnarray}
for the corrections in first order in $V$.
This leads to
\begin{eqnarray}
 \hat P_{1}^{1} &=& - {i\over \pi} {\ln (\beta U'/ 2\pi) \over 1-f(\epsilon)}
        \cos{\varphi \over 2} \hat P_{2}^{1}
\\
 \hat P_{2}^{1} &=& -{i\over 2} eV {f'(\epsilon)\over 1+f(\epsilon)} \cdot
        {C\over 1-f(\epsilon)} \sin (\varphi/2)
\end{eqnarray}
with the factor $C$ as defined in Eq.~(\ref{C}).

\section{Alternative derivation of Eq.~(4.7)}
\label{append_alternative}

As shown in Ref.~\onlinecite{Meir92} the current can be written as
\begin{equation}
   I = {e\over h} \sum_\sigma \int d\omega \, {\rm \bf tr} \left\{ {\bf G}^a
        {\bf \Gamma}^R {\bf G}^r {\bf \Gamma}^L \right\}
        \left( f_L - f_R \right)
\end{equation}
In this case for the linear conductance only equilibrium Green's functions are
involved.
These can be determined exactly up to all orders, e.g., by using an
equations-of-motions approach.
The result for the retarded Green's function is
\begin{equation}
   G^r(\omega) =
        \left( \begin{array}{cc}
        \omega - \epsilon + i \Gamma / 2
        & i (\Gamma / 2) \cos (\varphi / 2)
        \\
        i (\Gamma / 2) \cos (\varphi / 2)
        & \omega - \epsilon + i \Gamma / 2
        \end{array} \right) ^{-1}
\end{equation}
and the advanced Green's function is the complex conjugate.
As a consequence the transmission is
\begin{eqnarray}
   T(\omega) &=& {\Gamma^2 (\omega-\epsilon)^2 \cos^2(\varphi/2) \over
        (\omega-\epsilon)^2 + (\Gamma/2)^2(1+\cos(\varphi/2))^2 }
\nonumber \\
        &\times& {1\over (\omega-\epsilon)^2 +
        (\Gamma/2)^2 (1-\cos(\varphi/2))^2 } \, .
\end{eqnarray}
An expansion up to first and second order in $\Gamma$ yields
\begin{equation}
   T^{(1)}(\omega) = \pi \Gamma \delta(\omega-\epsilon)
        \cos^2(\varphi/2)
\end{equation}
for the first order, which proves Eq.~(\ref{two dots noninteracting}), and
\begin{equation}
   T^{(2)}(\omega) = {\rm Re} \, {\Gamma^2 \cos^2(\varphi/2) \over
        (\omega-\epsilon + i0^+ )^2}
\end{equation}
for the second order (cotunneling).
Note that for $\varphi = 0,\pm 2\pi, \pm 4\pi,\ldots$, the transmission in
second order is twice the sum of the transmissions through the dots taken
apart, as expected for constructive interference, while in first order
(at resonance) no factor 2 is involved.

\end{appendix}


\begin{references}
\bibitem{note1}
This is justified so long as the temperature is high enough to render the
Kondo physics irrelevant.
\bibitem{renormalization}
The renormalization terms derived in second order, ``cotunneling at
resonance'',\cite{Koenig_cotunneling} are negligible in this regime.
\bibitem{comment_reg}
The regularization $+i0^+$ is put here by hand, but can be derived within a
complete cotunneling theory.\cite{Koenig_cotunneling}
\bibitem{note2}
Here, we have chosen $t_L t_R \tilde t e^{- i\varphi}$ to be positive.
If we chose a minus sign, all our conclusions would be modified in a trivial
way, $\varphi \rightarrow \varphi + \pi$.
This would modify, though, details of the evolution of the transmission phase
with the gate voltage, which is not considered here.
\bibitem{note3}
This can be easily proven within a diagrammatic technique as introduced in 
Appendix~\ref{append_one_dot}.
The diagram of each contribution to the Green's function $G^{(1)}$ contains one 
``double vertex'' describing direct tunneling through the reference arm, say 
from the left to the right lead.
By interchanging the labels $L$ and $R$ for the tunnel lines connected to the
double vertex we can construct another diagram which contributes to $G^{(1)}$ as
well.
The double vertex now describes tunneling from right to left.
Since in equilibrium the left and right leads are described by the same Fermi 
function, the analytic expressions for the new diagram are almost identical to
the old ones.
The only difference is a reversal of the AB phase, 
$\varphi \rightarrow - \varphi$, due to the exchange of the double vertex.
This proves the desired symmetry relation.
It is straightforward to generalize this proof for Green's function $G^{(n)}$ of
arbitrary high order in $t^{\rm ref}$ and, thus, for the total Green's function
as well.
\bibitem{note4}
We consider here the generic case where the signs of all tunneling matrix
elements $t_{L, \rm dot 1}$, $t_{L, \rm dot 2}$, $t_{R, \rm dot 1}$, and
$t_{R, \rm dot 2}$ are the same (in the absence of flux).
Another choice of the relative phases among them will modify the detailed
conclusions of the following analysis.
\bibitem{note6}
For completeness we mention that further minus signs can appear in diagrams 
which include states with two or more fermions.
They are due to the fact that a two-vertex contribution such as
$c_\uparrow c^\dagger_\downarrow | \uparrow \rangle$
leads to a minus sign, $ - | \downarrow \rangle$ 
(see Refs.~\onlinecite{Koenig96,Koenig99}).
\end{references}
\end{document}